\documentclass[aps,prx,reprint,showpacs,showkeys,noeprint,longbibliography,superscriptaddress]{revtex4-2}
\usepackage{graphicx}
\usepackage{amsmath}
\usepackage{array}
\usepackage{amsfonts}
\usepackage{float}
\usepackage{xcolor}
\usepackage{dsfont}
\usepackage{calligra}
\usepackage{url}
\usepackage{appendix}
\usepackage{wasysym}
\usepackage{slashed}

\usepackage{cmap}
\usepackage[T1]{fontenc}
\usepackage{textcomp}
\usepackage{amsmath}
\usepackage{latexsym}
\usepackage{float}
\usepackage{amssymb}
\usepackage{graphicx}
\usepackage{hyperref}
\usepackage{xcolor}
\usepackage{url}
\usepackage{booktabs,tabularx,dcolumn}

\def\ba{\begin{eqnarray}}
	\def\ea{\end{eqnarray}}
\def\be{\begin{equation}}
	\def\ee{\end{equation}}
\def\bm{\begin{math}}
	\def\me{\end{math}}

\newcommand{\dummy}

\makeatletter
\newcommand{\fmarki}{*}
\newcommand{\fmarkii}{\ensuremath{\dagger}}
\newcommand{\fmarkiii}{\ensuremath{\ddagger}}
\newcommand{\fmarkiv}{\ensuremath{\mathsection}}
\newcommand{\fmarkv}{\ensuremath{\mathparagraph}}
\newcommand{\fmarkvi}{\ensuremath{\|}}
\newcommand{\fmarkvii}{**}
\newcommand{\fmarkviii}{\ensuremath{\dagger\dagger}}
\newcommand{\fmarkix}{\ensuremath{\ddagger\ddagger}}

\def\@fnsymbol#1{{\ifcase#1\or \fmarki\or \fmarkii\or \fmarkiii\or \fmarkiv\or \fmarkv\or \fmarkvi\or \fmarkvii\or \fmarkviii\or \fmarkix \else\@ctrerr\fi}}
\makeatother

\begin{document}
	\title{Dynamical crossovers and correlations in a harmonic chain of active particles}

	\author{Subhajit Paul}\email[]{ spaul@physics.du.ac.in, subhajit.paul@icts.res.in}
	\affiliation{International Center for Theoretical Sciences, 
		Tata Institute of Fundamental Research, 
		Bangalore-$560089$, India}
	\affiliation{Department of Physics and Astrophysics, University of Delhi, Delhi-$110007$, India}
	\author{Abhishek Dhar}\email[]{ abhishek.dhar@icts.res.in}
	\affiliation{International Center for Theoretical Sciences, 
		Tata Institute of Fundamental Research, 
		Bangalore-$560089$, India}
	\author{Debasish Chaudhuri} \email[]{debc@iopb.res.in}
	\affiliation{Institute of Physics, 
		Sachivalaya Marg, Bhubaneswar-$751005$, India}
	\affiliation{Homi Bhabha National Institute, Anushaktinagar, Mumbai-$400094$, India}
	
	\date{\today}

\begin{abstract}
We explore the dynamics of a tracer in an active particle harmonic chain, investigating the influence of interactions. Our analysis involves calculating mean-squared displacements (MSD) and space-time correlations through Green's function techniques and numerical simulations. Depending on chain characteristics, i.e., different time scales determined by interaction stiffness and persistence of activity, tagged-particle MSD exhibit ballistic, diffusive, and single-file diffusion (SFD) scaling over time, with crossovers explained by our analytic expressions. Our results reveal transitions in bulk particle displacement distributions from an early-time bimodal to late-time Gaussian, passing through regimes of unimodal distributions with finite support and negative excess kurtosis and longer-tailed distributions with positive excess kurtosis. The distributions exhibit data collapse, aligning with ballistic, diffusive, and SFD scaling in the appropriate time regimes. However, at much longer times, the distributions become Gaussian. Finally, we derive expressions for steady-state static and dynamic two-point displacement correlations, consistent with simulations and converging to equilibrium results for small persistence. Additionally, the two-time stretch correlation extends to longer separation at later times, while the autocorrelation for the bulk particle shows diffusive scaling beyond the persistence time. 
\end{abstract}
	
\maketitle
\section{Introduction}
Studies related to active particles have gained significant interest during the last decade due to their many interesting behaviors not shown by passive particles (see reviews  \cite{chate_2008,ramaswamy2010,romanczuk_2012, marchetti_13,beching_rmp_16} and references therein). These elements can self-propel and stay out of equilibrium by consuming energy from the external environment or dissipating their internal energy locally. They violate the condition of detailed balance and the equilibrium fluctuation-dissipation relation \cite{loi_08,fodor_16}. The direction of self-propulsion can undergo diffusive dynamics leading to a persistent motion of the element. Examples of active systems abound in nature, from molecular motors, cytoskeletons, cells, tissues, animals, and birds~\cite{Astumian2002, Berg1972, Niwa1994, Ginelli2015, Devereux2021}.  Although relatively fewer in number, artificial active systems have been developed over the past two decades utilizing different phoretic mechanisms like diffusiophoresis, thermophoresis, electrophoresis, etc., in active colloids~\cite{marchetti_13,beching_rmp_16, Illien2017, Paxton2004, Bricard2013, Bricard2015}, and vibrated asymmetric granular matter like vibrated rods, vibrobots, active spinners~\cite{Dauchot2019, Scholz2018, Narayan2007, Kudrolli2008, Deseigne2010, Farhadi2018}. They can perform collective motion, showing various self-organized complex patterns across different length scales \cite{beching_rmp_16, marchetti_13}. Typical examples are the flocking of birds, growth of bacterial colonies, the phoretic motion of synthetic Janus colloids, conformational changes of active polymeric objects in biological cells, etc.  In the presence of volume exclusion, persistent active  particles display a motility-induced phase separation (MIPS) \cite{stenhamm2013,cates_mips_15}.
\par
Three related models are often used in literature to describe the dynamics of active particles: (i) run-and-tumble particle (RTP) \cite{tailleur_rtp_08}, (ii) active Brownian particle (ABP) \cite{romanczuk_2012} and (iii) active Ornstein-Uhlenbeck particle (AOUP) \cite{fodor_16}.  
Due to their non-Markovian nature, associated with a long persistent time of the direction of self-propulsion, 
these particles show distinct non-Boltzmann distributions in the stationary state~\cite{basu2018,solon2015,Dhar2019,Patel2023}. 
A significant advance in the exact analytic understanding of the dynamics of individual active particles~\cite{Sevilla2014, Grossmann2016,Kurzthaler2018a,basu2018,malakar_18,Basu2019, Dhar2019,Shee2020,ion2020,Majumdar2020,Malakar2020,Basu2020, ion2022,Santra2021,chaudhuri2021,Patel2023,Shee2022,Shee2022a, Dean2021} has been achieved over the last few years. 
The prevalent theoretical method to study many-body active systems has been the phenomenological hydrodynamic approach~\cite{tailleur_rtp_08,marchetti_13,cates_mips_15,stenhamm2013}. Relatively little is known in terms of exact analytic results for microscopic models of interacting active particles, apart from some recent results~\cite{slowman_2017,das2020gap,satya2019,slowman2016,houssene2018,put2019,dolai_20,singh_2021,banerjee_22,touzo_2023,agranov2023}.
\par 
One remarkable dynamical behavior emerges for the dynamics of interacting particles moving in narrow channels so that the particles cannot cross each other maintaining their initial order. In this case, the mean-squared-displacement (MSD) of a tagged particle at large times grows as  $\sqrt{t}$ known as single-file diffusion (SFD), a behavior starkly different from the linear scaling observed in simple diffusion \cite{harris1965,kollman2003,lizana2008,hegde2014,krapivsky2015}. This behavior can be realized by probing the motion of a single particle in a collection of colloidal particles in one-dimensional channels~\cite{wei2000,lutz2004}.  While a comprehensive theoretical understanding of SFD in systems of passive Brownian particles exists, despite some recent attempts~\cite{singh_2021,banerjee_22,dolai_20}, a similar level of understanding is still lacking for active particle systems.
\par 
In this paper, we consider a linear chain of overdamped RTPs interacting harmonically with nearest neighbors.  
Recent calculations using Fourier transform in position space obtained  MSD~\cite{put2019} and correlation functions~\cite{singh_2021} in the RTP chain as series expansions in normal modes and demonstrated interesting dynamical crossovers.  For example, Ref.\cite{put2019} showed a crossover from ballistic to SFD via a transient regime of super-diffusion.  In this paper, we take a different approach, Fourier transforming the equations of motion in time and utilizing Green's function technique to obtain several exact and closed-form expressions. We find several dynamical crossovers in MSD that differ in detail from Ref.\cite{put2019}. The crossovers between ballistic, diffusive, and SFD in our model depend on competition between active persistence and bond stiffness. Our exact analytic calculations agree with numerical simulation results and help us to find the crossover times.
Moreover, we obtain crossovers in displacement distributions of a bulk particle from non-Gaussian to Gaussian behavior associated with the different dynamical regimes. Their change in characteristic is further quantified in terms of excess kurtosis in displacement. Finally, we obtain two-point and two-time correlation functions of displacement and stretching, with the stretch correlation showing late-time power-law decay due to length conservation.      
	
\par 
The rest of the paper is organized as follows. In Sec.~II, we discuss the model and present a summary of our main results. In Sec.~III, we present the analytical calculations. In Sec.~IV, the results are compared against direct numerical simulations of our model. Finally, in Sec.~V, we conclude with an outlook.

\section{Model, observables, and summary of results}

{\bf Model:} Our model consists of $N$ RTPs connected by harmonic springs \cite{singh_2021}.  In this case, the overdamped equations of motion for the system are given by 
\begin{eqnarray}\label{main_harmonic_eqn}
\dot{x}_l&=&-k(2x_l-x_{l-1}-x_{l+1}) + v_0\sigma_l
\end{eqnarray}
for $l=1,2,\ldots,N$, where { a spring constant $K$ controls the relaxation rate $k=K/\gamma$} for displacement in  harmonic potential, $v_0$ is the active velocity  and $\sigma_l$ represents  active dichotomous noise for the $l$-th RTP switching between $\sigma_l=1$ and $-1$ with a rate $\alpha$. The relaxation time corresponding to the harmonic interaction is  $\tau_k=k^{-1}$.
The corresponding correlation is
\begin{equation}\label{active_noise_crl}
\langle \sigma_l(t_1)\sigma_m(t_2) \rangle = e^{-2\alpha |t_1-t_2|}\delta_{l,m}\,,
\end{equation}
where $\tau_{\alpha}=1/\alpha$ controls the persistence time of the RTP. 
We consider two kinds of boundary conditions for the chain: {(i) Fixed boundaries}: $x_{N+1}=x_0=0$  and
(ii) Periodic boundary conditions (PBC): $x_{0}=x_N$ and $x_{N+1}=x_1$. 
We note that in the presence of PBC, the chain remains free to diffuse, and as a result, the center of mass of the chain shows asymptotic diffusion. However, for fixed boundaries, with boundary particles trapped by harmonic potentials, the position of the center of mass reaches a steady state in a time-scale $\gtrsim N^2 /k$. 
	
\begin{figure*}[t!]
		\centering
		\hskip -0.5cm
		\includegraphics*[width=17.50cm,height=5.0cm]{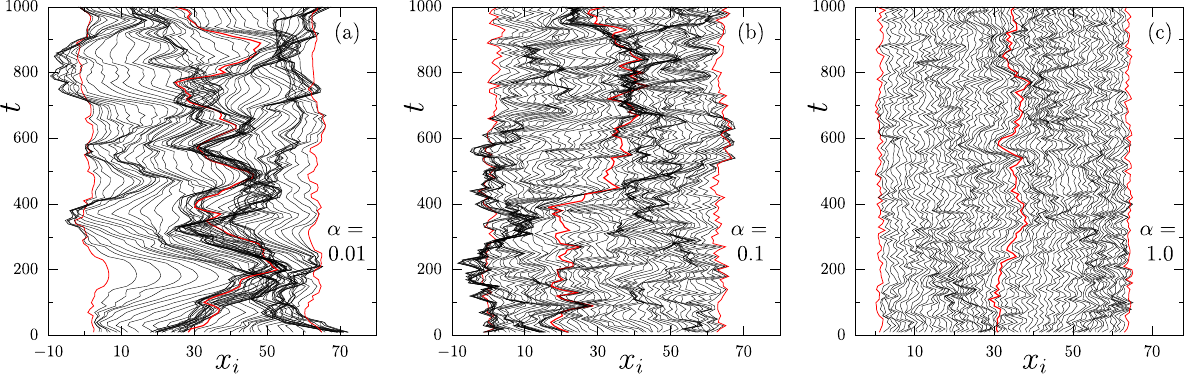}
		\caption{\label{fig_rtp_trajec} Plot of the space-time trajectories of $N=64$ harmonically coupled RTPs  for the tumbling parameter (a) $\alpha=0.01$, (b) $\alpha=0.1$ and (c) $\alpha=1.0$. In all the plots, the trajectories for bulk and the boundary particles are marked by red. For any particle (except at the boundaries), one sees more persistent motion with the value of $\alpha$ decreasing.} 
\end{figure*}

The equations of motion can be directly integrated using the Euler-Maclaurin scheme using $\delta t=10^{-3}$. In Fig.~\ref{fig_rtp_trajec}, we illustrate the typical dynamics of the system by showing the trajectories of the particles for two different values of $\alpha=0.01$ and $1.0$ for a chain of length $N=64$. The trajectories for bulk and the boundary particles are marked in red. The trajectories can cross each other as the interaction is a harmonic one. An important difference to notice is that the direction of any particle changes more frequently for $\alpha=1.0$ compared to $\alpha=0.01$, as $\alpha$ determines the persistent time $\tau_{\alpha}= \alpha^{-1}$. Due to this, the bulk particle for $\alpha=0.01$ can traverse a longer distance than the other one. However, the boundary particles can not move much due to their local confinement. Our main aim is to demonstrate the effects of this persistent motion in a system with single file diffusion on observables such as the MSD of tagged particles and on spatio-temporal correlations.  We now define the main observables of interest.

{\bf Observables:} The first quantity of interest is  the mean-squared-displacement (MSD) of a tagged RTP. Starting from $t_0$, the MSD for the $l$-th RTP, is defined as
\begin{equation}
\Delta_l(t)=\lim_{t_0 \to \infty} \left \langle [x_l(t+t_0) -x_l(t_0)]^2 \right\rangle\,,
\end{equation}
where $\langle ...\rangle$ denotes an average over different realizations, and we consider $t_0 \to \infty$ such that the results are independent of initial condition. 
For the chain with PBC, the MSD must be the same for all the particles due to translational invariance. However, with fixed boundaries, we expect the nature of $\Delta_l(t)$ to depend on the choice of $l$ with significant differences between bulk and boundary particles. 
	
For the case of fixed boundaries, we also study the evolution of the displacement distribution of the middle particle $P(\delta x_{N/2},t)$, with $\delta x_{N/2}(t)=x_{N/2}(t)-x_{N/2}(0)$. 
	
We also calculate the equal time and two-time correlations between displacements of different RTPs in the steady state. The general  two-time  correlation between displacements of $l$-th and $m$-th particles is given by
\begin{equation}\label{x_gen_correl}
C^x_{l,m}(t)= \lim_{t_0 \to \infty} \Big[\langle x_l(t_0) x_m(t_0+t) \rangle - \langle x_l(t_0)\rangle \langle  x_m(t_0+t) \rangle\Big]\,.
\end{equation}
By symmetry, the mean displacement $\langle x_l(t) \rangle=0$ and thus, the second term in the above equation does not contribute. 
	
A related quantity is a correlation between a local extension or the stretch variable    
$y_l(t)=x_{l+1}(t)-x_l(t)$, which is inversely proportional to local density. For a chain of length $L$, the sum total of all stretches must vanish. Thus $y_l(t)$ evolves obeying a continuity equation,
\begin{align}\label{eqn_y_current}
\Dot{y}_l = {\nabla}_l \cdot (-k \nabla_l y_l + v_0 \sigma_l)
\end{align}
where $\nabla_l$ denotes discrete differential operator such that $\nabla_l f_l = (f_{l+1}-f_l)$, and the conserved stochastic current $j_l = -k \nabla_l y_l + v_0 \sigma_l$.  
	The correlation is defined as 
	\begin{equation}\label{y_gen_correl}
		C^y_{l,m}(t)= 
		\Big[\langle y_l(t_0) y_m(t_0+t) \rangle - \langle y_l(t_0)\rangle \langle  y_m(t_0+t)\Big]\,
	\end{equation}
	where an average over $t_0$ is implemented. 
	We also consider the forms of the equal time correlations
	\begin{align}\label{main_xystat_correl}
		S^x_{l,m}=C^x_{l,m}(t=0),~~ S^y_{l,m}=C^y_{l,m}(t=0).
	\end{align}
	
	At this stage, a few comments are in order. 
	In Eqs.~\eqref{x_gen_correl}, \eqref{y_gen_correl} and \eqref{main_xystat_correl}, the second terms should vanish by symmetry. Indeed, they do not contribute to the correlations in numerical calculations when averaged over many independent initial configurations.
	For   PBC, all correlation functions depend only on the relative separation $|m-l|$, while for fixed boundaries, the correlations depend on both $m$ and $l$.
	The impact of activity remains prevalent on these quantities at short times and slowly disappears as time progresses. 
	
	\begin{table}[t!]
		\caption{Table summarises the behavior of the bulk particle MSD ($\Delta_{N/2}(t)$), its displacement distribution $P(\delta x_{N/2},t)$ and corresponding kurtosis $\kappa$ in different time regimes, for   $\alpha=50.0$ and $k=0.05$. The corresponding relaxation times are  $\tau_{\alpha}=0.02$ and $\tau_k=20.0$. }\label{tab1}
		\centering
		\begin{tabular}{|c|c|c|c|}
			\hline
			~~time~~&~$t \ll \tau_{\alpha}, \tau_k$~~&~~~$\tau_{\alpha} < t < \tau_k$~~&~~$t \gg \tau_{\alpha}, \tau_k$\\
			~~~&~~~~&~~~~&~~\\
			\hline
			~$\Delta_{N/2}(t)$~~&~~\text{ballistic}~~&~~~~diffusive 
			~~&~~SFD \\
			\hline
			~~~~&~non-Gaussian with~&~~~~&~~~~\\
			~~P($\delta x_{N/2},t$)~~&~~~two delta-function~&~~Gaussian~~&~Gaussian~\\
			~~~&~~~peaks at $\pm v_0 t$~&~~~~&~~\\
			\hline
			~~~&~~~~&~~~~&~~\\
			~~$\kappa$~~&~~negative ~~&~~0~~&~~0~~\\
			\hline
		\end{tabular}
	\end{table}	
	
	{\bf Summary of results:}
	The following are the main contributions of this paper:
	\begin{enumerate}
		\item We obtain analytic expressions for MSD of individual particles. We obtain closed-form expressions for bulk particles to describe crossovers between ballistic, diffusive, and single-file-diffusion (SFD) scaling. For a finite chain with pinned boundaries, MSD saturates earlier and to smaller values for particles near the boundary.

		\item  At short times, the displacement distributions of the central (bulk) particle show different characteristic features depending on persistence $\alpha^{-1}$ and interaction strength $k$, e.g., a bimodal distribution typical of free RTPs, unimodal but non-Gaussian distributions with finite support (negative kurtosis), and distributions with extended tails longer than Gaussian (positive kurtosis). However, at late times $t \gg [\tau_k, \tau_{\alpha}]$, all the distributions become Gaussian~(See Tables I and II). We find clear and separate data collapses for the distributions corresponding to different scaling regimes, e.g., ballistic, diffusive, and SFD.  
		
		\item The equal-time correlation functions  $S^x_{l,m}$ and $S^y_{l,m}$ show departures from equilibrium over a separation $v_0/\alpha$. In the limit of large $\alpha$, the results agree with equilibrium prediction.

		\item The two-time and two-point correlation of local stretch spreads over larger distances for longer time gaps. The same point correlation $C^y_{N/2,N/2}(t)$ remains unchanged over a time-scale $\tau_{\alpha}$, and decays in an approximate diffusive manner $ C^y_{N/2,N/2}(t) \sim t^{-1/2}$ for longer times.

	\end{enumerate}

	\begin{table}[htb]
		\caption{Same as Table~\ref{tab1} but for $\alpha=0.01$ and $k=1.0$. Here the two time scales are $\tau_{\alpha}=100.0$ and $\tau_k=1.0$}\label{tab2}
		\centering
		\begin{tabular}{|c|c|c|c|}
			\hline
			time~~&~$t \ll \tau_{\alpha}, \tau_k$~~&~~~$\tau_k < t < \tau_{\alpha}$~~&~~$t \gg \tau_{\alpha}, \tau_k$\\
			~~~&~~~~&~~~~&~~\\
			\hline
			~$\Delta_{N/2}(t)$~~&~~\text{ballistic}~~&~~~~ballistic 
			~~&~~SFD \\
			\hline
			P($\delta x_{N/2},t$)~~&~non-Gaussian with~&~~approaching~~&~~~~\\
			~~~~&~~~heavy tails ~&~~Gaussian~~&~Gaussian~\\
			\hline
			~~&~~~~&~~~~&~~~~\\
			~$\kappa$~~&~~positive ~~&~~positive~~&~~0~~\\
			\hline
		\end{tabular}
	\end{table}

	{\bf Mapping to ABP and AOUP:}	The equation of motion for ABP evolves as,
	\begin{eqnarray}\label{abp}
		\dot{x}_l&=&-k(2x_l-x_{l-1}-x_{l+1}) + v_0\cos\theta_l \nonumber\\
		\dot \theta_l &=& \sqrt{2 D_r} \eta_l(t)
	\end{eqnarray}
	where $\eta_l(t)$ is a Gaussian white noise with $\langle \eta_l(t)\rangle=0$ and $\langle \eta_l(t) \eta_m(0)\rangle=\delta_{l,m}\delta(t)$. In this case, the correlation function of active velocity is determined by that of the heading directions $\langle \cos(\theta_l-\theta_m)\rangle = e^{-D_r t}$, where $D_r$ is the rotational diffusion constant. Thus using  $D_r=2\alpha$ maps the results for RTP to that of ABP, up to the second moment.	\\
	The AOUP evolves as,
	\begin{eqnarray}\label{aoup}
		\dot{x}_l&=&-k(2x_l-x_{l-1}-x_{l+1}) + v_l \nonumber\\
		\dot v_l &=& -\gamma_v v_l + \sqrt{2 D_v} \eta_l(t)
	\end{eqnarray}
	with the Gaussian noise in active velocity obeying $\langle \eta_l(t)\rangle=0$ and $\langle \eta_l(t) \eta_m(0)\rangle=\delta_{l,m}\delta(t)$. In this case the correlation between active velocities $\langle v_l(t) v_m(0) \rangle = \langle v_l^2(0) \rangle e^{-\gamma_v t}\, \delta_{lm}$ with $\langle v_l^2(0) \rangle = D_v/\gamma_v$, where $D_v$ is the diffusion constant and $\gamma_v^{-1}$ is the relaxation time related to memory. Thus, up to the second moment, the results of RTP can be mapped to AOUP if $\gamma_v=2\alpha$ and $D_v/\gamma_v=v_0^2$. 

	\section{Analytic results}
	
	Here, we describe Green's function-based approach which is particularly suitable for obtaining late-time properties~\cite{dhar2008,ion2022}. 
	We start by writing the equation of motion  Eq.~\eqref{main_harmonic_eqn} in the matrix form
	\begin{equation}\label{eom_pinned_matrix}
		\dot{{X}}=-{{{\Phi}} {X}} +{v}_0{{\Sigma}}\, ,
	\end{equation}
	where ${{X}}=({x}_1,{x}_2, \dots , {x}_N)^T$, and the matrix ${{\Phi}}$ is a symmetric tri-diagonal matrix with  elements: 
	\begin{equation}
		\Phi_{l,m}=k(2\delta_{l,m}-\delta_{l,m-1}-\delta_{l,m+1})\,,
	\end{equation}
	and ${{\Sigma}}=(\sigma_1,\sigma_2, ..., \sigma_N)^T$ is the vector containing the active noises. Fourier transforming Eq.~\eqref{eom_pinned_matrix}, we get:
	\begin{equation}
		\tilde{{X}}(\omega)=v_0{\mathcal{\Tilde{G}}(\omega)}~\tilde{{\Sigma}}(\omega)\,,
	\end{equation}
	where $\tilde{{X}}(\omega)=(\tilde{x}_1,\tilde{x}_2, ..., \tilde{x}_N)^T$ and $\tilde{{\Sigma}}(\omega)=(\tilde{\sigma}_1,\tilde{\sigma}_2, ..., \tilde{\sigma}_N)^T$ and  $\tilde{x}_l$ and $\tilde{\sigma}_l$ are the Fourier transforms of $x_l$ and $\sigma_l$, respectively, defined as
	\begin{equation}
		\tilde{x}_l(\omega)=\int_{-\infty}^{\infty} x_l(t)e^{-i\omega t} dt, ~~~~~ \tilde{\sigma}_l(\omega)=\int_{-\infty}^{\infty} \sigma_l(t)e^{-i\omega t} dt\,.
	\end{equation}
	The Green's function is defined as, 
	\begin{equation}{\mathcal{\Tilde{G}}}(\omega)=(i\omega\mathds{1}+{\Phi})^{-1},
	\end{equation}
	which is a $N\times N$ matrix with  $\mathds{1}$ being the identity matrix.

	\begin{widetext}
		The two-time correlation $C_{l,m}^x(t)=\big\langle x_l(t)x_m(0)\big\rangle$ can be obtained as elements of the matrix
		\begin{eqnarray}\label{main_correl_t_matrix}
			C^x(t) &=& \langle X(t) X^T(0) \rangle \nonumber \\
			&=& \frac{v_0^2}{4\pi^2}\int_{-\infty}^{\infty} \int_{-\infty}^{\infty} \tilde{\mathcal{G}}(\omega_1) \left\langle \tilde{{\Sigma}}(\omega_1)\tilde{{\Sigma}}^T(\omega_2)\right\rangle \tilde{\mathcal{G}}^T(\omega_2) e^{i\omega_1 t} ~d\omega_1 ~d\omega_2\, .
		\end{eqnarray}
		Using the form of the noise correlation matrix $\left\langle \tilde{{\Sigma}}(\omega_1)\tilde{{\Sigma}}^T(\omega_2)\right\rangle$ in Fourier space (shown in Appendix \ref{append_noisecorrel_fourier}) the elements of $C^x(t)$ matrix can be written more explicitly as:
		\begin{equation}\label{eq_two_point_correl}
			C_{l,m}^x(t) =  \frac{2 v_0^2 \alpha}{\pi} \int_{-\infty}^{\infty} \frac{\sum_k \tilde{\mathcal{G}}_{l,k}(\omega)~ \tilde{\mathcal{G}}_{m,k}^{T}(-\omega)}{4\alpha^2+\omega^2} e^{i\omega t} ~d\omega
		\end{equation}
		As shown in Appendix \ref{append_eq_active_correl}, using the fact that  matrix $\Phi$ has a simple tri-diagonal structure, we can obtain two representations ($R_1$ and $R_2$) for the elements of the Green's function $\tilde{\mathcal{G}}(\omega) = (\Phi+ i\omega \mathds{1})^{-1}$ as:
		\begin{align}\label{main_glmeq_r1r2}
			\tilde{\mathcal{G}}_{l,m}(\omega) &= \sum_{s=1}^N \frac{\phi_s(l)\phi_s(m)}{\lambda_s+i \omega} ~~~~~~~~~~~~~~~~~~~~~~~~~~~~~~~~~~~~~~~{\rm~~---}~R_1 \\
			&= \frac{\sin (l q) \sin [(N-m+1)q]}{\sin q \sin [(N+1) q]}, ~~~~{\rm for }~~l \leq m {~~~~~~~\rm~~---}~R_2, \label{GijB}
		\end{align}
		where  $\lambda_s=2 \big[1-\cos (s \pi/(N+1))\big]$,  $\phi_s(l)=\sqrt{2/(N+1)} \sin(s l \pi/(N+1))$, $s=1,2,\ldots,N$, are the eigenvalues and eigenfunctions of the matrix $\Phi$, and $q=\cos^{-1} (1+i \omega/2)$.
		In Eq.~\eqref{GijB},  the matrix elements with $l>m$ are obtained using the fact that $\tilde{\cal  G}$ is a symmetric matrix~\cite{hu_1996}.
		These lead, respectively, to the following two more explicit forms for the correlations (for details, see Appendix~\ref{append_eq_active_correl}): 
		\begin{align}
			C_{l,m}^x(t)=& ~2 v_0^2 \alpha \sum_{s=1}^N  \frac{\phi_s(l) \phi_s(m)}{4 \alpha^2-\lambda_s^2} \left(\frac{e^{-\lambda_s t}}{\lambda_s}-\frac{e^{-2 \alpha t}}{2 \alpha}\right) {~~~~~~~~~\rm~~---}~R_1 \nonumber\\
			=\label{crl_t_rep1} & ~\frac{2 v_0^2 \alpha}{\pi} \int_{-\infty}^{\infty} d \omega \frac{e^{i \omega t}}{4\alpha^2+\omega^2} F_{l,m}(\omega)~{~~~~~~~~~~~~~~~~~~~\rm~~---}~R_2,
		\end{align}
		\begin{align}\label{main_ggomega_ij}
			{\rm where}~~F_{l,m}(\omega)&=-\frac{1}{\omega} \text{Im}\left[\frac{\sin (l q) \sin (N-m+1) q}
			{\sin (q) \sin \big((N+1) q\big)\big)}\right], ~~~~\text{for}~~ l \leq m\, .
		\end{align}
		From these we can obtain the static steady state correlation $S_{l,m}^x$ as:
		\begin{align}\label{main_sscrl_t0_eq}
			S_{l,m}^x&=C^x_{l,m}(t=0)= v_0^2 \sum_{s=1}^N  \frac{\phi_s(l) \phi_s(m)}{ \lambda_s   (2 \alpha+\lambda_s)}  {~~~~~~~~~~~~~\rm~~---}~R_1 \nonumber\\
			&= \frac{2 v_0^2 \alpha}{\pi} \int_{-\infty}^{\infty} \frac{F_{l,m}(\omega)}{4\alpha^2+\omega^2}~d\omega~~ ~~~~~~{\rm for}~~l\leq m~~~~~~ --- R_2~.
		\end{align}
		{The two-point correlation for the stretch variable $y$ can be written as
			\begin{align}
				C_{l,m}^y(t)=\langle y_l(0) y_m(t) \rangle =  \Big\langle \Big[x_{l+1}(0)-x_l(0)\Big] \Big[x_{m+1}(t)-x_m(t)\Big] \Big \rangle\, . 
				\label{eq_Cy}
			\end{align}
			From this $S_{l,m}^y$ can be obtained with $t=0$.
		}
		
		The tagged particle MSD can be obtained from  the following general two-time displacement correlation matrix  given by
		\begin{equation}
			\mathcal{C}_{l,m}(t) = \Big \langle \Big [x_l(t)-x_l(0)\Big ] \Big[x_m(t)-x_m(0)\Big] \Big \rangle, 
		\end{equation}
		which is then easily related to the  static and dynamic correlations defined above, namely to $S_{l,m}^x$ and $C_{l,m}^x(t)$, as 
		\begin{equation}
			\mathcal{C}_{l,m}(t)= 2 S_{l,m}(0)-C_{l,m}^x(t)-C_{m,l}^x(t)\,.
		\end{equation}
		The diagonal element with $m=l$ provides the MSD for $l$-th RTP in a general form as:
		\begin{equation}\label{msd_diag_matrix}
			\Delta_{l}(t)={\mathcal{C}}_{l,l}(t)=\frac{2 v_0^2 \alpha}{\pi} \int_{-\infty}^{\infty} \frac{\sum_k \tilde{\mathcal{G}}_{l,k}(\omega)~ \tilde{\mathcal{G}}_{l,k}^{T}(-\omega)}{4\alpha^2+\omega^2} \left(2-2\cos \omega t\right)~ d\omega\,.
		\end{equation}
	\end{widetext}
	
	For the case of a very long chain with fixed boundary conditions, we expect that most particles in the bulk will have the same MSD up to times before finite size effects show up. For a chain with periodic boundary conditions, translation symmetry implies that all particles have the same MSD for all times (for any $N$). In either case, this allows us to obtain a simpler form for the bulk MSD defined as
	\begin{align}\label{bulk_msd_full_form}
		\Delta(t)&=\frac{1}{N}\sum_{l=1}^N \Delta_l(t) \nonumber \\
		&=\frac{2 v_0^2 \alpha}{\pi} \int_{-\infty}^{\infty}  \mathcal{B}(\omega) \frac{(2-2 \cos \omega t)}{(4\alpha^2+\omega^2)} ~d\omega,
	\end{align}
	where 
	\begin{align}\label{main_Bomega_trace_bulk}
		\mathcal{B}(\omega)=\frac{1}{N} \sum_{l,m} \tilde{\mathcal{G}}_{l,m}(\omega) \tilde{\mathcal{G}}_{m,l}(-\omega)\nonumber =\frac{1}{N} {\rm{Tr}} \Big[\tilde{\mathcal{G}}_{l,m}(\omega) \tilde{\mathcal{G}}_{m,l}(-\omega)\Big]\,,\\
	\end{align}
	and $\rm{Tr}$ denotes the trace of the matrix. In  Appendix \ref{append_2}  we show that   $\mathcal{B}(\omega)$ has the simple form:
	\begin{equation}\label{main_Bomega_finalform}
		\mathcal{B}(\omega)= 2 ~\text{Re}\left[\frac{1}{2\omega \sqrt{\omega^2-4ik\omega}}\right]\,,
	\end{equation}
	as shown in Eq.~\eqref{app_Bomega_finalform}. This has a simple pole near $\omega=0$ and a branch cut. 
	Getting a closed-form expression from this integral is difficult, but evaluating this numerically is straightforward. Below, we present numerical simulation results and discuss analytic predictions for scaling behaviors in different $t$ regimes.

	\section{Comparisons with numerical simulations and asymptotic forms}
	
	\subsection{MSD and displacement distribution of a tagged particle in bulk}
	We consider a few different regimes in which we look at the scaling behaviors of $\Delta_{N/2}(t)$. Depending upon the time scale, we can classify the behavior of the active particles in the following regimes:
	
	(i) short time: $t \ll \tau_{\alpha}, \tau_k \rightarrow$ free, ballistic
	
	(ii) intermediate time:
	\begin{enumerate}
		\item $\tau_{\alpha}  \ll t \ll \tau_k \rightarrow$ free, diffusive,
		\item $\tau_k  \ll t \ll \tau_{\alpha} \rightarrow$ interacting, ballistic,
	\end{enumerate}
	(iii) long time: $t \gg \tau_{\alpha},\tau_k \rightarrow$ single-file diffusion.\\
	
	
	\begin{figure}[t!]
		\centering
		\includegraphics*[width=8.0cm,height=7.10cm]{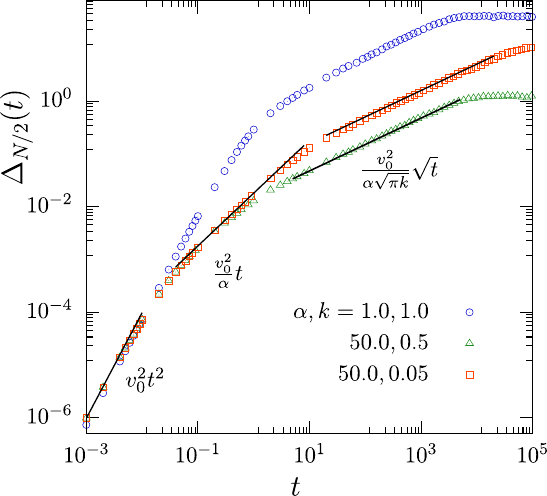}
		\caption{\label{msd_mid_alpha_k} Plots of MSD of the bulk RTP $\Delta_{N/2}(t)$ vs $t$ for different choices of $\alpha$ and $k$ parameters, as mentioned in the figure, for  a chain with $N=128$.  For $\alpha=50.0$ and $k=0.05$, one can see different scaling regimes better due to nice separation of two timescales $\tau_{\alpha}$ and $\tau_k$. The solid lines through the data correspond to the different power-law behaviors $t^2$, $t$ and $t^{1/2}$ with their corresponding amplitudes given by Eqs.~\eqref{ampli_ballistic_short}, \eqref{ampli_diff_inter} and \eqref{ampli_subdiff_late}, respectively. For all the data, we have used the active velocity $v_0=1.0$. }
	\end{figure}
	
	\begin{figure*}[t!]
		\centering
		\hskip -0.2cm
		\includegraphics*[width=18.0cm,height=7.70cm]{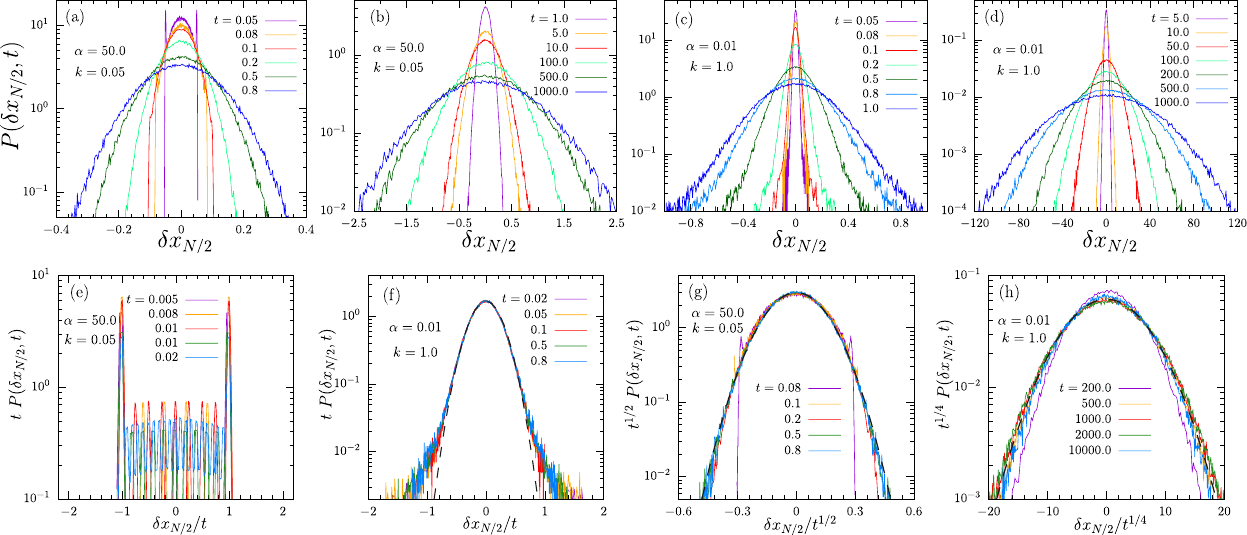}
		\caption{\label{fig_bulk_flucdist_scld} (a)-(d)~Plots of the time evolution of the distributions  $P(\delta x_{N/2},t)$ of a  bulk RTP for two sets of activity parameter $\alpha$ and interaction $k$. (e)-(h)~Data collapse at various dynamical regimes showing the scaling form $P(\delta x_{N/2}, t)=t^{-\mu} f_P(\delta x_{N/2}/t^\mu)$ with $\mu=1$ in ballistic, $\mu=1/2$ in diffusive, and $\mu=1/4$ in SFD regime. More explicitly, (e) and (f) corresponds to the ballistic scaling ($t < \tau_{\alpha}, \tau_k$), whereas in (g) ($\tau_{\alpha} < t < \tau_k$) and (h) ($t > \tau_{\alpha}, \tau_k$) we show the corresponding diffusive and sub-diffusive scalings, respectively. In (f)-(h) the dashed lines correspond to the Gaussian fits. Whereas we see nice Gaussian behavior during diffusive as well as single-file scaling, there is strong deviation from Gaussianity during the  ballistic motion at early times.  The  $\delta$-function peaks in (a) and (e),  near $\pm v_0 t$, correspond to the particle moving ballistically without switching its direction.}
	\end{figure*}

	\par
	Below, we discuss the scaling properties observed in the above-mentioned regimes in detail:
	
	(i)~For  $t \ll \tau_{\alpha}, \tau_k$, the interaction is unimportant and the particles perform fully persistent motion.  In this case, the leading contribution to the integral in Eq.~\eqref{bulk_msd_full_form} comes for modes $\omega \gg k$ and we get $\mathcal{B}(\omega) \approx 1/\omega^2$. 
	The integral simplifies to
	
	\begin{eqnarray}\label{ampli_ballistic_short}
		\Delta_{N/2}(t) &\approx& \frac{2v_0^2\alpha }{\pi} \int_{-\infty}^{\infty} \frac{4 \sin^2\left(\frac{\omega t}{2}\right)}{\omega^2(4\alpha^2+\omega^2)}~d\omega\nonumber\\
		&=& v_0^2 t^2 \left( 1+{\cal O}(\alpha t)\right)\,,
	\end{eqnarray}
	which corresponds to a short-time ballistic behavior of {\em free} RTPs.\\
	
	(ii)-($a$)~In the intermediate time regime, $\tau_{\alpha} \ll t \ll \tau_k$, we can use $\omega \gg k$, which leads to ${\cal B}(\omega) \approx 1/\omega^2$.  Since in this regime $\alpha \gg \omega$, approximating $(4 \alpha^2 +\omega^2)\approx 4\alpha^2$ the integration of Eq.~\eqref{bulk_msd_full_form} can be expressed as
	\begin{equation}\label{ampli_diff_inter}
		\Delta_{N/2}(t) 
		\approx \frac{v_0^2 }{2\pi \alpha} \int_{-\infty}^{\infty} \frac{ \sin^2(\frac{\omega t}{2})}{\omega^2} ~d\omega = 2 D_{\text{eff}}t\,,
	\end{equation}
	which shows a diffusive scaling with effective diffusion constant $D_{\text{eff}}={v_0^2}/{2\alpha}$.

	\par 
	(ii)-($b$)~We can consider another intermediate regime with small $\alpha$ and large $k$, i.e.,  {$\tau_k \ll t \ll \tau_{\alpha}$}. In this limit $\mathcal{B}(\omega)$ can be approximated as 
	${\cal B}(\omega) \approx \Big[2^{3/2} k^{1/2} \omega^{3/2} \Big]^{-1}$ and
	the MSD from Eq.~\eqref{bulk_msd_full_form} as
	
	\begin{eqnarray}\label{ampli_ballistic_inter}
		\Delta_{N/2}(t)
		&\approx&  \frac{2\sqrt{2} v_0^2 \alpha }{\pi\sqrt{k}} \int_{-\infty}^{\infty} \frac{\sin^2\left(\frac{\omega t}{2}\right)}{\omega^{3/2}(4\alpha^2+\omega^2)}~d\omega \nonumber \\
		&=& 
		\left(\frac{\alpha}{2 k} \right)^{1/2}\, v_0^2 t^2  \,(1+{\cal O}(\alpha^2 t^2))
	\end{eqnarray}
	With $\alpha t \ll 1$, the leading order again shows ballistic $t^2$ scaling as in the short time limit but with a different prefactor that depends on both persistence and interaction.

	(iii)~We now show that asymptotically, for $t \gg \tau_{\alpha}, \tau_k$,  the bulk particle in an infinite chain shows a sub-diffusive behavior. At long times, only the lowest frequency  modes contribute, and  we approximate 
	$\mathcal{B}(\omega) \approx \Big[2^{3/2} k^{1/2} \omega^{3/2} \Big]^{-1}$ as in the last case. Moreover,  since $\alpha \gg \omega$, approximating $(4 \alpha^2 +\omega^2)\approx 4\alpha^2$ the integration can be performed to get, 
	\begin{eqnarray}\label{ampli_subdiff_late}
		\Delta_{N/2}(t)
		&\approx&  \frac{\sqrt{2} v_0^2 }{2\pi\alpha \sqrt{k}} \int_{-\infty}^{\infty} \frac{\sin^2\left(\frac{\omega t}{2}\right)}{\omega^{3/2}}~d\omega \nonumber \\
		&=& 2 D_{\rm eff} 
		(\pi k)^{-1/2}
		t^{1/2},~{\rm with~} 
		D_{\rm eff}=\frac{v_0^2}{2 \alpha}
	\end{eqnarray}
	being the active free particle diffusivity. This has the same form as that of a passive system of harmonically coupled Brownian particles for which ${\Delta}^{\text{eq}}_{{N/2}}(t)=2 {D} \sqrt{\gamma/(\pi K)}t^{1/2}$~\cite{Lizana2010}, where $D$ is the equilibrium diffusivity and relaxation rate $k=K/\gamma$. The only difference is, for active particles, the coefficient depends on activity via $D_{\rm eff}=\frac{v_0^2}{2 \alpha}$.

	
	
	Consider now the case studied in Refs.~\cite{dolai_20,banerjee_22}, of a set of RTP particles with short-ranged repulsive interactions. If we make the identification
	$(\rho v_0)^{-1}=\gamma/K$, which corresponds roughly to equating the collisional and relaxation time scales, then   Eq.~\eqref{ampli_subdiff_late} gives the following prediction for the long time MSD in the RTP gas with short-ranged repulsion: 
	\begin{equation}
		\Delta_{N/2}(t)= 2\frac{D_{\rm eff}}{\sqrt{\pi \rho v_0}} t^{1/2}, 
	\end{equation}
	which, apart from a factor of $2/\sqrt{\pi}$, agrees with the estimate in  Ref.~\cite{dolai_20}.

	{\bf Crossover times:} We can also obtain the crossover points by comparing the MSD expressions in different scaling regimes. The ballistic to diffusive crossover time $t_{1}^{c}$ is determined by comparing $v_0^2 (t_{1}^{c})^2 \approx 2 D_{\rm eff} (t_{1}^{c})$ leading to $t_{1}^{c} \sim 1/\alpha$. It is consistent with the original criterion of getting the intermediate regimen at $t > \tau_{\alpha}$.  
	The crossover time $t_{2}^c$ from intermediate time effective diffusion to the late time single-file-diffusion can be obtained by using $2D_{\text{eff}}(t_{2}^c) \approx 2(D_{\text{eff}}/\sqrt{\pi k})(t_{2}^c)^{1/2}$ to get $t_{2}^c \sim 1/(\pi k)$. 
	Finally, a possibility arises in which the initial ballistic regime directly crosses over to the SFD regime at $t_{3}^c$ such that $v_0^2 (t_{3}^c)^2 \approx 2(D_{\text{eff}}/\sqrt{\pi k})(t_{3}^c)^{1/2}$ to give $t_{3}^c \sim (\alpha \sqrt{\pi k})^{-2/3}$. The condition for this direct crossover is $t_{3}^c < t_{1}^c$.
	
	As we now discuss, the above estimates show reasonable comparison against direct numerical simulations.

	
	{\bf Comparison of MSD with numerical simulations:} In Fig.~\ref{msd_mid_alpha_k} we plot $\Delta_{N/2}(t)$ for different choices of $\alpha$ and $k$ for a chain of length $N=128$ with the boundary particles pinned locally by harmonic traps. Due to pinning, asymptotically, the MSD of the bulk particle saturates at time scales of order $\tau_{\text{sat}}\sim N^2/k$. Thus, for $k=0.05$, the saturation is barely visible in a plot up to $t=10^5$. Before saturation, all the plots show a short-time ballistic regime $\Delta_{N/2} \sim t^2$ and a  late-time SFD regime $\Delta_{N/2} \sim t^{1/2}$. 
	
	Plots for $(\alpha,k)=(50,0.5)$ and $(50,0.05)$ capture both the ballistic-diffusive and diffusive-SFD crossovers before the steady-state saturation, as for both of them $t_2^{c} > t_1^{c} $ and $t_3^{c} > t_1^{c}$. Moreover, as discussed before, to observe the intermediate regime of simple diffusion, the criterion $\tau_{\alpha} \ll t \ll \tau_k$ has to be obeyed for which having a clear separation of the two-time scales,  $\tau_{\alpha}$ and $\tau_k$ is important. Because of this, the diffusive regime is best observed over a significantly long period for $\alpha=50.0$ and $k=0.05$. 
	
	On the other hand, for $(\alpha,k)=(1.0,1.0)$,  $t_2^{c} < t_1^{c}$ and $t_3^{c} < t_1^{c}$;  thus, the intermediate diffusive regime disappears, and the MSD shows direct crossover from initial ballistic to late-time SFD behavior, before the steady state saturation. 
	
	The solid lines in Fig.~\ref{msd_mid_alpha_k} show plots of  Eqs.~\eqref{ampli_ballistic_short}, \eqref{ampli_diff_inter} and \eqref{ampli_subdiff_late} in the different scaling regimes, i.e., $t^2, t ~\text{and}~ t^{1/2}$, including the numerical values of the analytically obtained prefactors to show excellent agreement with the simulation results. 
	
	{\bf Displacement distributions of the bulk}:  Here we consider the full probability distributions $P(\delta x_{N/2},t)$ of the displacement of the central particle $\delta x_{N/2}(t)= x_{N/2}(t) - x_{N/2}(0)$.  In Fig.~\ref{fig_bulk_flucdist_scld}, we numerically characterize the time evolution of $P(\delta x_{N/2})$ and approximate data-collapse of distributions in different scaling regimes.
	In Figs.~\ref{fig_bulk_flucdist_scld}(a)-(d) we show the time evolution of normalized $P(\delta x_{N/2},t)$ for two combinations of $\alpha$ and $k$. In (a) and (c), we plot $P(\delta x_{N/2},t)$ for short times, whereas (b) and (d) show the corresponding distributions at late times. They are plotted separately since the ranges of both $P$ and $\delta x_{N/2}$ get very different with $P(\delta x_{N/2})$ spreading over time. 
	
	Note that the time scales corresponding to the chosen  parameters are $\tau_{\alpha}, \tau_k= (0.02, ~20.0) ~\text{and} ~(100.0,~ 1.0)$. 
	The choice of parameters ensures that in the first case (Figs.~\ref{fig_bulk_flucdist_scld}(a),(b)) the interaction affects the dynamics much later than the persistence time governing the ballistic behavior, as $\tau_k \gg \tau_{\alpha} $. 
	Whereas, for the second choice of parameters (Figs.~\ref{fig_bulk_flucdist_scld}(c),(d)) interaction affects even in the ballistic regime as  $\tau_k \ll \tau_{\alpha} $.  
	The distributions at short times (i.e., in (a) and (c)) show qualitatively different natures depending on the parameter choices, the distributions at late times look qualitatively similar in both cases (see Figs.~\ref{fig_bulk_flucdist_scld}(b) and (d)).   
	
	In the first case, with $\alpha=50.0, k=0.05$, at very short times, i.e., $\tau_{\alpha} \lesssim t\ll \tau_k$,  corresponding to ballistic MSD, in addition to the central Gaussian-like maximum due to flipping of heading direction ($t=0.05 > \tau_\alpha$), the distribution has two delta function-like peaks, approximately at $\pm v_0 t$. Similar behavior was observed before for non-interacting persistent walkers in Ref.~\cite{dhar2001, malakar_18}. Here this behavior appears even in the presence of interaction, but at $t \ll \tau_k$ the time required for the interaction to influence individual dynamics. In this regime the distributions have the scaling form $t^{-1}f_s(\delta x_{N/2}/t)$, where $f_s(y)$ has a strongly non-Gaussian shape, see  Fig.~\ref{fig_bulk_flucdist_scld}(e). 
	
	For longer times, the height of the two peaks at $\pm v_0 t$ starts to decrease and eventually, the distribution becomes unimodal. For times, 
	$\tau_{\alpha} < t < \tau_k$ 
	the particle shows a diffusive behavior (as obserevd in Fig.~\ref{msd_mid_alpha_k}) and thus the distributions follow the scaling $t^{-1/2}f_g(\delta x_{N/2}/t^{1/2})$, where $f_g(y)$ is a Gaussian in Fig.~\ref{fig_bulk_flucdist_scld}(g). 
	
	Now for the second choice of parameters, i.e., $\alpha=0.01, k=1.0$, the dynamics at short times are still ballistic, and the distributions follow the ballistic scaling form $t^{-1}f_s(\delta x_{N/2}/t)$ as shown in Fig.~\ref{fig_bulk_flucdist_scld}(f) with $f_s(y)$ is non-Gaussian. 
	However, for $t \gg \tau_{\alpha}, \tau_k$ we observe the scaling form  $t^{-1/4}f_g(\delta x_{N/2}/t^{1/4})$ corresponding to SFD as interaction dominates. Further, $f_g(y)$ has a Gaussian form as shown in Fig.~\ref{fig_bulk_flucdist_scld}(h). Such an SFD scaling is also observed for the late time distributions of Fig.~\ref{fig_bulk_flucdist_scld}(b)~(not shown). In Fig.~\ref{fig_bulk_flucdist_scld}(f)-(h), the black dashed lines show the Gaussian fits. 
	In the diffusive and SFD regimes Fig.~\ref{fig_bulk_flucdist_scld}(g),(h), the agreement with Gaussian is good; however, in the ballistic regime, although it shows a nice fit for small $\delta x_{N/2}$, a clear deviation with longer tails are observed in Fig.~\ref{fig_bulk_flucdist_scld}(f).  Such deviations from Gaussian can be quantified via the kurtosis of the distributions. This we show in the following. 
	
	

{\bf Kurtosis of displacement:}
The departures of the distributions from the Gaussian nature can be quantified using the excess kurtosis 
\begin{equation}\label{kurto_defn}
	\kappa(t)= \frac{\left \langle \left[\delta x_{N/2}(t) - \mu(t) \right]^4\right \rangle}{\left \langle \left[\delta x_{N/2}(t) - \mu(t) \right]^2\right \rangle^2} -3.0\,,
\end{equation}
where $\mu(t)=\langle \delta x_{N/2}(t) \rangle$ is the mean of $\delta x_{N/2}(t)$ and $\langle \dots \rangle$ defines average over different initial conditions. By definition, $\kappa=0$ for a Gaussian distribution. 

\begin{figure}[t!]
	\centering
	\hskip -0.4cm
	\includegraphics*[width=8.80cm,height=7.10cm]{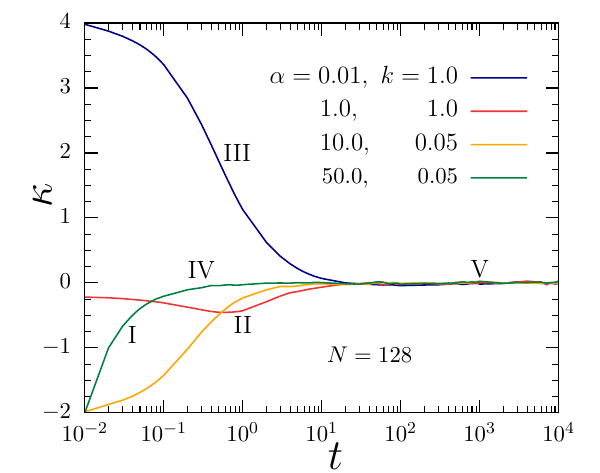}
	\caption{\label{flucdist_alpha_time} Evolution of excess kurtosis $\kappa (t)$ corresponding to the displacement of the bulk particle $\delta x_{N/2}$ for parameter values $(\alpha,k)$ denoted in the figure legend.  
	}
	\label{fig_Pdx}
\end{figure}

In Fig.~\ref{fig_Pdx} we plot $\kappa$ versus $t$ for a few different combinations of $\alpha$ and $k$. We see that depending upon the values of $\alpha$ and $k$, i.e., the time-scales $\tau_{\alpha}$ and $\tau_k$ the early time behavior is markedly different which is also seen from $P(\delta x_{N/2},t)$ in Fig.~\ref{fig_bulk_flucdist_scld}. However, at $t \gg \tau_{\alpha}, \tau_k$, $\kappa=0$ corresponds to an SFD Gaussian. Note that $\kappa$ remains positive for $\alpha=0.01$ and $k=1.0$, before vanishing at $t \gtrsim \tau_{\alpha}$. Whereas for the other graphs, we see the kurtosis varies from negative to zero. 

For better insight, we identify different points marked by (I)-(V) for which the distributions are shown in Fig.~\ref{fig_bulk_flucdist_scld}. At (I), the negative value of $\kappa$ is determined by the two delta function peaks in Fig.~\ref{fig_bulk_flucdist_scld}(e). (II) also corresponds to a negative $\kappa$, but for unimodal distributions with finite support (less extreme outliers than Gaussian) at $t \lesssim (\tau_{\alpha}, \tau_k)$ (not shown). (III) marks a region of positive $\kappa$ with unimodal distributions having longer tails approaching zero slower than Gaussian~(Fig.~\ref{fig_bulk_flucdist_scld}\,(f)\,). Here $t \ll \tau_{\alpha}$ and $t \sim  \tau_k$, a ballistic but reasonably interacting regime. 
(IV) corresponds to an intermediate diffusive regime  for which $P(\delta x_{N/2},t)$ fits to a Gaussian~(Fig.~\ref{fig_bulk_flucdist_scld}(g)\,). Finally, (V) corresponds to the SFD regime, common for all combinations of $\alpha$ and $k$ for $t \gg \tau_{\alpha}, \tau_k$ in which the distributions become Gaussian due to interactions~(Fig.~\ref{fig_bulk_flucdist_scld}(h)\,).


{\bf Nature of displacement for bulk particle:}
{
	Using MSD, distributions, and kurtosis, we arrive at the following conclusion regarding the changing nature of displacement distribution. At $t \ll \tau_{\alpha}$, ballistic motion is associated with a large positive deviation from Gaussian in terms of a bimodal distribution captured by the positive kurtosis. At the ballistic diffusive crossover, which is essentially a single particle effect, the kurtosis becomes negative, and the displacement distribution has a finite support meaning a sharper localization than Gaussian. This happens if the persistence is not too high, $\alpha > 0.01$. For a longer time, interaction couples fluctuations along the chain, making this distribution Gaussian due to the central limit theorem. In this regime, interaction leads the MSD to display single-file diffusion, with the displacements following a Gaussian distribution. At later times, the whole chain can show simple diffusion if the boundaries are not locally pinned.}

\begin{figure}[t!]
	\centering
	\hskip -0.5cm
	\includegraphics*[width=9.10cm,height=7.10cm]{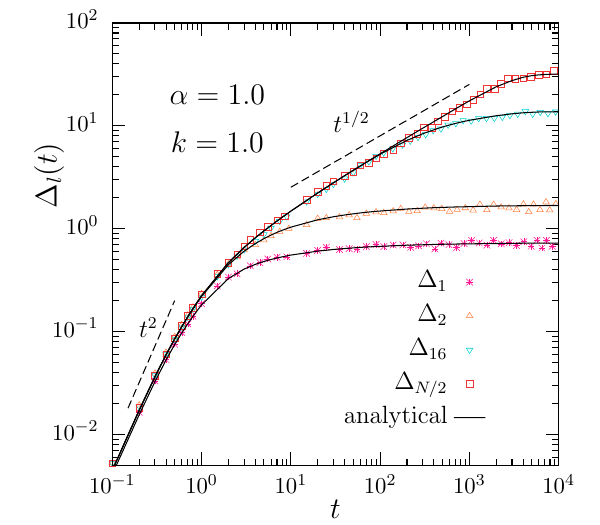}
	\caption{\label{msd_jtag}  Plot of mean-squared-displacements ($\Delta_l(t)$) for the first ($l=1$), second ($l=2$), $16$-th ($l=16$), and the middle ($l=N/2$) particle, for a chain of $N=128$. The continuous solid lines through the data correspond to the analytical forms provided in Eq.~\eqref{msd_diag_matrix}. The dashed lines correspond to the power-law behaviors showing ballistic $t^2$ and SFD $t^{1/2}$ scaling.}
\end{figure}

\subsection{MSD for any tagged RTP} 
\par
In the presence of boundary pinning, with the boundary particles locally trapped in external harmonic potentials, the dynamics of a tagged particle change as one goes from bulk toward the boundary of the chain. 
A direct numerical evaluation of Eq.~\eqref{msd_diag_matrix} can be used to obtain the MSD of $l$-th particle $\Delta_l$. 
Using left-right symmetry around the center $l=N/2$, the calculation of $\Delta_l$ in one side of the central particle suffices to describe the local behavior of the whole chain.

To gain a better analytic insight into the calculation of $\Delta_l(t)$, from Eq.~\eqref{msd_diag_matrix} we consider $\tilde{G}_{l,l}(\omega)$ defined as
\begin{equation}
	\tilde{G}_{l,l}(\omega)=\sum_{k=1}^{N} \tilde{\mathcal{G}}_{l,k}(\omega) \tilde{\mathcal{G}}^T_{l,k}(-\omega)=\sum_{k=1}^{N} |\tilde{\mathcal{G}}_{l,k}(\omega)|^2\,,
	\label{eq_Hij}
\end{equation}
using $\tilde{\mathcal{G}}_{l,k}^T(\omega)=\tilde{\mathcal{G}}_{l,k}(\omega)$ in the last step,
as the inverse of the symmetric matrix $\mathcal{A}=(\Phi+i\omega \mathds{1})$ is also symmetric and 
$\tilde{\mathcal{G}}_{l,k}(-\omega)=\tilde{\mathcal{G}}_{l,k}^*(\omega)$.
Given the tri-diagonal nature of the matrix, for $l \le m$ we can write~\cite{hu_1996} 
\begin{equation}
	\mathcal{\Tilde{G}}_{l,m}(\omega)=\frac{\mathcal{D}_{1,l-1} \mathcal{D}_{m+1,N}}{\mathcal{D}_{1,N}}
	\label{eq_g_Dij}
\end{equation}
where $\mathcal{D}_{i,j}$ is the determinant of the sub-matrix of $\mathcal{A}$ starting with $i$-th row and column and ending with $j$-th row and column. 

\par


Using $(2+i\omega)=2 \cos q$ with a complex quantity $q$, the determinant $\mathcal{D}_{l,m}$ in terms of $q$ can be expressed as $\mathcal{D}_{l,m}= \sin [(m-l+2)q]/\sin q$. 
This leads to the simplification, 
\begin{equation}
	\mathcal{\Tilde{G}}_{l,m}(\omega)= \frac{\sin(lq) ~\sin[(N-m+1)q]}{\sin q~\sin[(N+1)q]}\,~~~~\text{for}~~l \leq m\,.
	\label{eq_Gq2}
\end{equation}
For the symmetric matrix, to get  $\tilde{\mathcal{G}}_{l,m}(\omega)$ for $l>m$, the position of $l$ and $m$ in the above expression has to be swapped. Further to get the hermitian conjugate $\mathcal{\Tilde{G}}^T_{l,m}(-\omega)$, we replace $q$ by $q^*$ in Eq.~\eqref{eq_Gq2}. The details of the above calculation are shown in Appendix \ref{gmatrix_ij}.\\ 




The above form can be used to write, 
\begin{eqnarray}
	\tilde{G}_{l,l}(\omega) &=&\sum_{k=1}^{N} |\tilde{\mathcal{G}}_{l,k}(\omega)|^2 \nonumber \\ 
	&=& \sum_{k=1}^{l} |\tilde{\mathcal{G}}_{l,k}(\omega)|^2 + \sum_{k=l+1}^{N} |\tilde{\mathcal{G}}_{l,k}(\omega)|^2 \nonumber \\
	&=& \sum_{k=1}^{l} \frac{|\sin(kq)|^2 |\sin(N-l+1)q|^2}{|\sin q|^2 |\sin(N+1)q|^2} \nonumber \\
	&+& \sum_{k=l+1}^{N} \frac{|\sin(lq)|^2 |\sin(N-k+1)q|^2}{|\sin q|^2 |\sin(N+1)q|^2}\, .
\end{eqnarray}
	%
%
%
A much simpler closed-form expression can be found for the bulk particle $l=N/2$ in a {\em large enough} chain, using $(N/2+1)\approx N/2$ to obtain



\begin{align}
	&\tilde G_{\frac{N}{2} \frac{N}{2}}(\omega) = \frac{|\sin (Nq)|^2}{\omega|\sin q|^2 |\sin (N+1)q|^2} \bigg[\omega |\sin (Nq)|^2 \nonumber \\ 
	&\left. + {\rm Re}\left\{ i\sqrt{\omega(4i+\omega)}\cos\left(\frac{Nq^*}{2}\right)\sin\left(\frac{Nq}{2}\right) \right\} \right]\, .
\end{align}

\par
In Fig.~\ref{msd_jtag}, we illustrate the difference in the dynamics between bulk and boundary elements by plotting the MSDs $\Delta_l(t)$ for a few different values of $l$ from boundary ($l=1$) to bulk ($l=N/2$) for a chain of $N=128$ for parameters $\alpha=1.0$ and $k=1.0$. The data points are from simulations, and the solid lines show direct numerical integration results of Eq.~\eqref{msd_diag_matrix}. They show excellent agreement.   
These plots display direct crossover from ballistic to SFD due to the particular choice of $\alpha$ and $k$ that do not allow time-scale separation and, as a result, ballistic-diffusive crossovers. 
The main conclusion from these plots is regarding the clear difference between the bulk and boundary dynamics. While at the bulk locations, $l=N/2$, MSD displays a late-time ballistic to SFD crossover before saturation, MSD at locations near the boundary feels the pinning effect much earlier. Their MSD saturates beyond the ballistic regime, even before getting the opportunity to show any SFD behavior.

\begin{figure}[t!]
	\centering
	\hskip -0.4cm
	\includegraphics*[width=8.4cm,height=7.30cm]{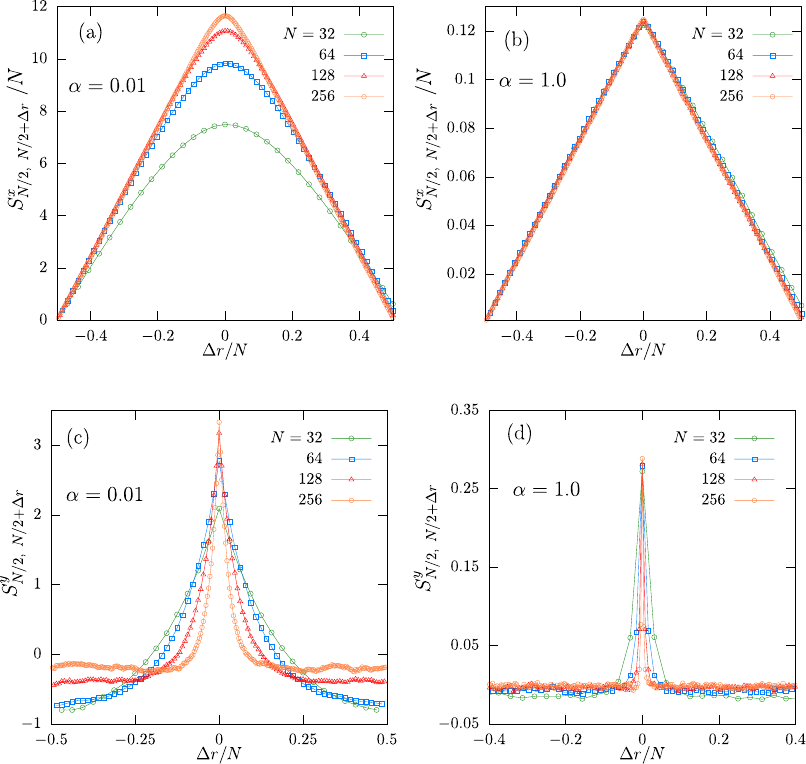}
	\caption{\label{crl_alpha_N} (a)-(b) Simulation results for scaled correlation $S_{l,m}^x/N$ versus the scaled distance $\Delta r/N \equiv (m-N/2)/N$ for different values of $N$ for two values of $\alpha=0.01$ and $1.0$. (c)-(d) We plot correlations for the  stretch variable $y_l$, i.e., $S^y_{l,m}$ versus $\Delta r/N$ for the corresponding  same values of $\alpha$ and chain lengths $N$. In all the plots we fix $l=N/2$ and $m=N/2+\Delta r$ where $\Delta r$ varies from $-N/2$ to $N/2$. We used the spring constant $k=1$ and active speed $v_0=1$ in these plots.}
\end{figure}

\begin{figure*}[t!]
	\centering
	\hskip -0.5 cm
	\includegraphics*[width=18.00cm,height=5.00cm]{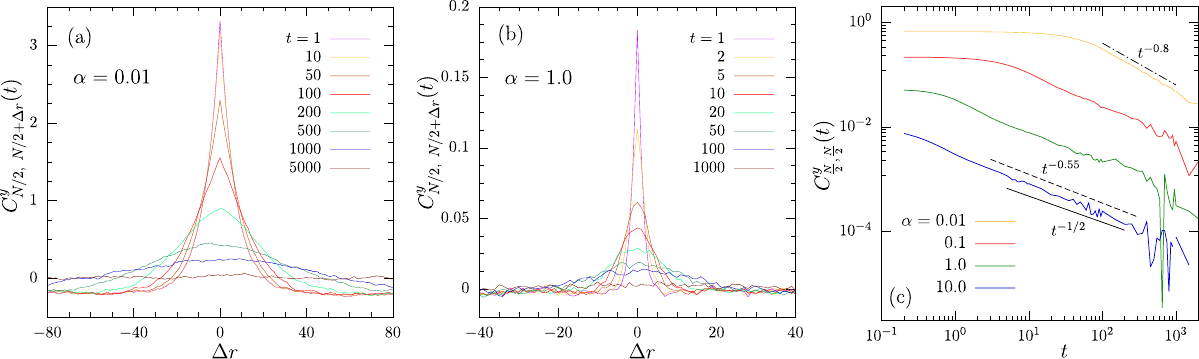}
	\caption{\label{ymax_t_alpha} (a) Two-time correlations $C_{N/2,N/2+\Delta r}^y(t)$ of the stretch parameter $y$  versus the distance $\Delta r=m-N/2$ between two particles for different times $t$ as mentioned for $\alpha=0.01$.  (b) Similar to (a), but for $\alpha=1.0$. (c) Plot shows the evolution of the maximum value of $C^y_{l,m}$, i.e., auto-correlation $C_{N/2,N/2}^y(t)$ (with $\Delta r=0$) for the bulk particle versus $t$  for different values of $\alpha$. The power-law lines for $\alpha=0.01$ and $10.0$ with exponents $\sim 0.8$ and $\sim 0.55$. All data presented here are for a chain of length $N=256$. 
	}
\end{figure*}

\subsection{Equal time two-point correlations}
In an interacting system, displacements at different positions and times are correlated. 
The steady-state two-point correlation between any two particles $l$ and $m$ is given by the two representations in Eq.~\eqref{main_sscrl_t0_eq} with $F_{l,m}(\omega)$ defined by Eq.~\eqref{main_ggomega_ij}.
Notably, the steady-state correlation is independent of the coupling constant $k$. 
A representative heat map of this correlation is presented in Appendix~\ref{append_ss_t0_correlmatrix}, showing how the correlation decreases as one goes from bulk to the boundary.
The expression in Eq.~\eqref{main_sscrl_t0_eq} simplifies in the large $\alpha$ limit to
\begin{align}\label{main_eq_ss_xcrlij_R1R2}
	S^{x(\text{eq})}_{l,m}&=D_{\rm{eff}} \sum_{s=1}^N  \frac{\phi_s(l) \phi_s(m)}{ \lambda_s }
	= \frac{D_{\rm{eff}}}{\pi} \int_{-\infty}^{\infty} F_{l,m}(\omega)~d\omega  \,,
\end{align}
where $v_0^2/2\alpha=D_{\rm{eff}}$. 
In the large $\alpha$ (small persistence) limit, the system quickly {\em equilibrates} to an effective temperature $T_{\rm eff}$ given by $D_{\rm{eff}}= k_B T_{\rm eff}/\gamma$. Thus we recover the equilibrium expression, see Eq.~\eqref{eq_app_static_crl_phiinv}.
More explicitly, 
%
\begin{align}\label{eqlbrm_Sx_N_scaling}
	S^{x(\text{eq})}_{l,m} = D_{\rm{eff}} \Phi^{-1}_{l,m} = \frac{D_{\rm{eff}}}{k} \frac{l(N-m+1)}{(N+1)}\,, ~~\text{for} ~~l \le m.
\end{align}
where we have used $\Phi_{l,m}^{-1} = \tilde{\mathcal{G}}_{l,m}(\omega=0) = \frac{l(N-m+1)}{(N+1)}$. For $l >m$, the indices in the final expression of Eq.~\eqref{eqlbrm_Sx_N_scaling} must be swapped.


Setting $l=N/2$ and $m=N/2+\Delta r$ we get 
%
\begin{align}
	S_{l,m}^{x\text{(eq)}} = D_{\rm{eff}} \frac{N}{4}\left(1-2 \frac{|\Delta r|}{N} \right).
	\label{eqlbrm_crl_Nscaling}
\end{align}

Similar to $S_{l,m}^x$ we define the equal time correlation for the `stretch' variable $y_l=x_{l+1}-x_l$ as well. Following Eq.~\eqref{y_gen_correl} we define $S^y_{l,m}$ for the steady state as 
\begin{align}
	S^y_{l,m}&= \lim_{t_0 \to \infty} \langle y_l(t_0) y_m(t_0) \rangle \nonumber\\ 
	&=\lim_{t_0 \to \infty} \left\langle \left(x_{l+1}(t_0)-x_l(t_0)) (x_{m+1}(t_0)-x_m(t_0)\right)\right\rangle\,,\nonumber
\end{align}
as $\langle y_l(t_0)\rangle = 0$ by symmetry.

In Figs.~\ref{crl_alpha_N}(a) and (b) we plot steady-state scaled correlation functions $S^x_{l,m}/N$ choosing $l=N/2$  as a function of scaled separation $\Delta r/N=(l-m)/N$ for two different values of $\alpha=0.01$ and $1.0$. As shown above, at equilibrium, the correlation is expected to decrease linearly with separation. Fig.~\ref{crl_alpha_N}(a) shows that the departure from equilibrium vanishes at a longer separation; for longer chains, the fractional distance at which the departure is visible gets smaller. At higher $\alpha$ (Figs.~\ref{crl_alpha_N}(b)\,) an excellent equilibrium-like data-collapse in agreement with Eq.~\eqref{eqlbrm_crl_Nscaling} is observed.  On the other hand, for small $\alpha$ ($\alpha=0.01$), such a data collapse and the cusp in correlation near $\Delta r=0$ disappears. Note that the expression in Eq.~\eqref{main_sscrl_t0_eq} and the numerical simulation results for the correlation show excellent agreement~(see Fig.~\ref{appfig_heatmap_t0_crl}(b) in Appendix \ref{append_ss_t0_correlmatrix}).


In Figs.~\ref{crl_alpha_N}(c) and (d) we show the corresponding steady-state correlation functions for stretch $S^y_{l,m}$ vs $\Delta r/N$ for the same choices of the parameter values as in  Figs.~\ref{crl_alpha_N}(a) and (b). Defined in terms of the discrete first-order derivative of $x_l$ the two-point correlation for $y_l$, the cusp at $\Delta r=0$ in $S^x_{N/2,N/2+\Delta r}$ translates into Dirac delta-function at $\Delta r=0$ in $S^y_{N/2,N/2+\Delta r}$. We observe such behavior and approximate data collapse at $\alpha=1.0$ in Fig.~\ref{crl_alpha_N}(d). However, this expectation is invalid for large persistence, as seen from the correlation functions plotted for $\alpha=0.01$ in Fig.~\ref{crl_alpha_N}(c).
Since the total stretch $\sum_{l=1}^{N-1} y_l$ is constant for a chain of constant length, the large positive correlation at small separations leads to anti-correlation at longer separations. 

\subsection{Two-point two-time correlation}
 The two-time correlation of displacements $C^x_{l,m}(t)$ at two different points along the chain is given by the two representations of Eq.~\eqref{crl_t_rep1}. Appendix \ref{app_Cx} shows a plot for this quantity's time evolution.  
	
Here, we focus on such a correlation for the local stretch variable $y_l$. Note that an increase in stretch is equivalent to a decrease in local density. Also, as clear from Figs.~\ref{crl_alpha_N}(c)-(d), this quantity does not have a dependence on $N$ in the steady state. The two-time correlation is defined in Eq.~\eqref{eq_Cy}.

In Figs.~\ref{ymax_t_alpha}(a)-(b) we show $C^y_{l,m}(t)$ with $l=N/2$ and $m=N/2+\Delta r$ as a function of separation $\Delta r$ with increasing $t$. We use a chain of $N=256$. The correlation spreads over longer separation with time. For $\alpha=1.0$ the starting profile is more delta-function-like compared to the one for $\alpha=0.01$. 
The initial anti-correlation at larger values of $\Delta r$ decreases, and the profile becomes flatter along the chain at later times. Even though, at late times, they show an approximately  Gaussian profile (as $y_l$ satisfies a diffusion-like equation Eq.~\eqref{eqn_y_current}), one observes significant deviations in the profiles at early times. In Fig.~\ref{ymax_t_alpha}(c)  we plot the correlation at $\Delta r=0$, i.e., the auto-correlation $C^y_{N/2,N/2}$ vs $t$. It should decay as $\sim t^{-1/2}$ for a diffusive process. At time scales $t > \alpha^{-1}$, one expects this diffusive scaling which is approximately realized for $\alpha \gtrsim 0.1$. The finite size effect might mar the deviations seen at smaller $\alpha$.  
		

		\section{Conclusion}
		We have studied the dynamics and correlations of a tracer in a linear chain of active RTPs with nearest-neighbor harmonic interactions. We considered the boundary particles to be in local traps, i.e., their positions do not fluctuate much. Such  boundary conditions help the system to reach a non-equilibrium steady state. Activity is modeled via a dichotomous random noise which switches its direction with a rate $\alpha$. This is the only adjustable parameter which sets the persistence length and thus by varying $\alpha$ we move from the active to the passive limit which corresponds to large value of $\alpha$.

		There are three time-scales present in the system, i.e., the persistence time $\tau_{\alpha}$,  the interaction time-scale $\tau_k$ and the system-size dependent time-scale $\tau_N \sim N^2/k$ at which the boundary effects appear. Using the Fourier transform method and by solving the equations of motion using the appropriate Green's function,  we have obtained a closed-form expression for the MSD of a tracer in bulk. From this, the behavior of MSD in different time limits has been extracted. With a nice separation between $\tau_{\alpha}$ and $\tau_k$ with $\tau_k \gg \tau_{\alpha}$ we find a  crossover from ballistic to diffusive dynamics in the intermediate time regime. However, for any values of $\alpha$ and $k$, if $t \gg \tau_{\alpha}, \tau_k$, the MSD shows an SFD-like sub-diffusive behavior. Apart from the scaling, We obtained an exact expression of the SFD {\em diffusivity}, i.e., the prefactor $2 D_{\rm eff}/\sqrt{\pi \rho v_0}$ showing its dependence on density, active velocity, and persistence through $D_{\rm eff} = v_0^2/2\alpha$. Even though, this SFD behavior is generic in such an interacting system, for particles close to the boundary this can not be observed as the finite size effect appears early and the MSD saturates. Differences in the dynamics of a tagged particle from bulk to the boundary have been observed from simulations as well as from our analytic results.  Our detailed analysis shows that at early times the full distribution for the bulk has non-Gaussian behavior which is a signature of the active nature, i.e., the persistence length of the particle. Depending upon $\tau_{\alpha}$ and $\tau_k$ the non-Gaussian behaviors have different forms. However, at late times, for $t \gg \tau_{\alpha}, \tau_k$ the distributions become Gaussian with an SFD scaling $t^{1/4}$ appearing due to its interaction with other particles.
		
		Finally, we obtained analytic expressions for displacement and stretch correlations that agree with numerical simulation results and show strongly non-equilibrium behavior that persists over shorter separation and time gaps. At longer separation or time they decorrelate in an equilibrium-like or diffusive fashion.

		Our analytic results for RTP up to the second moment can be readily mapped to that of ABP and AOUP models. The particular predictions we made regarding MSD, displacement distributions, and correlations are amenable to direct experimental observations, e.g., using channel-confined active colloids or using shape-asymmetric vibrated granular particles in narrow channels having transverse dimensions smaller than twice the particle size.
		
		\section*{Conflicts of interest}
		There are no conflicts to declare.
		
		\begin{acknowledgements}
			S.P. acknowledges ICTS-TIFR, DAE, Govt. of India under project no. RTI4001 for a research fellowship.   D.C. acknowledges research grants from DAE (1603/2/2020/IoP/R\&D-II/150288) and SERB, India (MTR/2019/000750), and thanks ICTS, Bangalore for an Associateship and support during the meeting ``Active Matter and Beyond" (code: ICTS/AMAB2024/11).  AD  thanks Urna Basu for useful discussions and  acknowledges support from the Department of Atomic Energy, Government of India, under Project No. RTI4001.
		\end{acknowledgements}
		
		\onecolumngrid
		~~~~~
		\appendix
		
		\section{Noise correlation in Fourier space $\langle \tilde{\sigma}(\omega_1) \tilde{\sigma}(\omega_2)\rangle $}\label{append_noisecorrel_fourier}
		The correlation $\langle \tilde{\sigma}(\omega_1) \tilde{\sigma}(\omega_2)\rangle $ for a RTP is:
		\begin{align}
			\langle \tilde{\sigma}(\omega_1) \tilde{\sigma}(\omega_2) \rangle \nonumber =	\int_{-\infty}^{\infty} \int_{-\infty}^{\infty} \langle {\sigma}(t_1) {\sigma}(t_2)\rangle~ e^{-i\omega_1 t_1}e^{-i\omega_2t_2} ~dt_1~ dt_2
		\end{align}
		with $\langle \sigma(t_1)\sigma(t_2)\rangle= e^{-2\alpha|t_1-t_2|}$ as mentioned in Eq.~\eqref{active_noise_crl}. 
		Now, splitting $|t_1-t_2|$ for $t_1 > t_2$ and $t_2 >t_1$ we obtain the following two terms as:
		\begin{align}
			\langle \tilde{\sigma}(\omega_1) \tilde{\sigma}(\omega_2) \rangle \nonumber =\int_{-\infty}^{\infty}e^{-i\omega_1 t_1} dt_1 \int_{-\infty}^{t_1} e^{-2\alpha(t_1-t_2)} e^{-i\omega_2t_2} dt_2 \nonumber 
			+ \int_{-\infty}^{\infty}e^{-i\omega_2 t_2} dt_2 \int_{-\infty}^{t_2} e^{-2\alpha(t_2-t_1)} e^{-i\omega_1 t_1} dt_1
		\end{align}
		
		This calculation provides us:
		\begin{equation}
			\langle \tilde{\sigma}(\omega_1) \tilde{\sigma}(\omega_2) \rangle \\
			=\frac{2\pi (4\alpha-i\omega_1-i\omega_2) \delta(\omega_1+\omega_2)}{(2\alpha-i\omega_1)(2\alpha-i\omega_2)}\,,
		\end{equation}
		which is used in Eq.~\eqref{main_correl_t_matrix}.

\section{Two-point Correlations} \label{append_eq_active_correl}
The two-point, two-time correlation function as written in Eq.~\eqref{eq_two_point_correl} for the active particles is defined as 
\begin{equation}\label{twotime_xcrl_matrixform}
C_{l,m}^x(t) =  \frac{2 v_0^2 \alpha}{\pi} \int_{-\infty}^{\infty} \frac{\sum_k \tilde{\mathcal{G}}_{l,k}(\omega)~ \tilde{\mathcal{G}}_{m,k}^{T}(-\omega)}{4\alpha^2+\omega^2} e^{i\omega t} ~d\omega
\end{equation}
From this, the static or equal-time correlation can be expressed as, i.e., with $t=0$
\begin{equation}\label{stat_xcrl_matrixform}
S_{l,m}^{x} = \langle x_l(0) x_m(0) \rangle = \frac{2 v_0^2 \alpha}{\pi} \int_{-\infty}^{\infty} \frac{\sum_k \tilde{\mathcal{G}}_{l,k}(\omega)~ \tilde{\mathcal{G}}_{m,k}^{T}(-\omega)}{4\alpha^2+\omega^2} ~d\omega
\end{equation}
First we show results for the equilibrium limit, i.e., for passive particles.  This limit is easily obtained using $\alpha \to \infty$ and $v_0 \to \infty$ keeping the diffusion constant $D=v_0^2/2\alpha$ constant. Thus, the equal-time equilibrium correlation for passive particles can be written as
\begin{equation}\label{eqcrl_from_act}
S_{l,m}^{x(\text{eq})} = \frac{D}{\pi} \int_{-\infty}^\infty d\omega\, \sum_k \tilde{\mathcal{G}}_{l,k}(\omega)~ \tilde{\mathcal{G}}_{m,k}^{T}(-\omega)  \equiv k_BT (\Phi^{-1})_{l,m}
\end{equation}
Below we describe two methods using which we can calculate $S_{l,m}^x$ or $C_{l,m}^x(t)$.\\
\textbf{Method I:}\\ 
The second expression in Eq.~\eqref{eqcrl_from_act} can be directly written using the Boltzmann distribution of displacements in equilibrium. Expanding in the eigenbasis of $\Phi$ we can write 
\begin{equation}\label{eq_app_static_crl_phiinv}
S_{l,m}^x= k_BT(\Phi^{-1})_{l,m} = k_BT\langle l|s \rangle \frac{1}{\lambda_s} \langle s|m \rangle = k_BT\sum_{s=1}^N \frac{\phi_s(l)~ \phi_s(m)}{\lambda_s}, ~~~~\text{where}~~~ \Phi |s\rangle = \lambda_s |s\rangle,
\end{equation}
where $\lambda_s$  is the eigenvalue corresponding to the eigenfunction $\phi_s(l)=\langle l |s \rangle$. In the above expression, we used $\gamma=1$. 
		
Now to use the Green's function $\Tilde{\mathcal{G}}_{l,m}(\omega)= ([\Phi+i\omega]^{-1})_{l,m}$ we can use the same eigenbasis of $\Phi$ with eigenvalues $\lambda_s+i\omega$ and we  get 
		\begin{equation}
			\Tilde{\mathcal{G}}_{l,m}(\omega)= \langle l|s \rangle \frac{1}{\lambda_s+i\omega} \langle s|m \rangle = \sum_{s=1}^N \frac{\phi_s(l)~ \phi_s(m)}{\lambda_s+i\omega}, ~~~~\text{where}~~~ (\Phi+i\omega \mathds{1}) |s\rangle = (\lambda_s+i\omega) |s\rangle
		\end{equation}
		Using this, we can calculate 
		\begin{equation}
			\sum_{k=1}^N \Tilde{\mathcal{G}}_{l,k}(\omega) \Tilde{\mathcal{G}}^T_{m,k}(-\omega) = \sum_{s,s'} \frac{\phi_s(l) ~\Big(\sum_{k} \phi_s(k)~ \phi_{s'}(k)\Big)~ \phi_{s'}(m)}{(\lambda_s+i\omega) (\lambda_{s'} -i\omega)}= \sum_{s=1}^N \frac{\phi_s(l)~ \phi_s(m)}{\lambda_s^2+\omega^2}\,,
		\end{equation}
		{where we used the completeness condition $\sum_{k=1}^N \phi_s(k)~ \phi_{s'}(k) = \delta_{s,s'}$.}
		With this, we write the equilibrium static correlation function as
		\begin{equation}\label{app_sij_eq_defn}
			S_{l,m}^{x(\text{eq})} = \frac{D}{\pi} \int_{-\infty}^{\infty} \sum_s \frac{\phi_s(l)~ \phi_s(m)}{\lambda_s^2+\omega^2} d\omega = D \sum_{s=1}^N \frac{\phi_s(l)~ \phi_s(m)}{\lambda_s} \equiv \frac{D}{\gamma} \sum_{s=1}^N \frac{\phi_s(l)~ \phi_s(m)}{\lambda_s}
		\end{equation}
		which is the identity expected in equilibrium, Eq.~\eqref{eq_app_static_crl_phiinv} with drag coefficient $\gamma=1$.
		
		\noindent\textbf{Method II:} \\   
		In this method, instead of expanding $\Tilde{\mathcal{G}}_{l,m}$ in its eigen basis, we write $\sum_k \tilde{\mathcal{G}}_{l,k}(\omega)~ \tilde{\mathcal{G}}_{m,k}^{T}(-\omega)$ as (depending upon our choices of $l$ and $m$)
		\begin{equation}
			\sum_{k=1}^N \tilde{\mathcal{G}}_{l,k}(\omega)~ \tilde{\mathcal{G}}_{m,k}^{T}(-\omega)
			=\sum_{k=1}^l \tilde{\mathcal{G}}_{k,l}(\omega)~ \tilde{\mathcal{G}}_{m,k}^{T}(-\omega)\nonumber 
			+\sum_{k=l+1}^m \tilde{\mathcal{G}}_{l,k}(\omega)~ \tilde{\mathcal{G}}_{k,m}^{T}(-\omega)\nonumber 
			+\sum_{k=m+1}^N \tilde{\mathcal{G}}_{l,k}(\omega)~ \tilde{\mathcal{G}}_{m,k}^{T}(-\omega)
		\end{equation}
		Writing $2+i\omega=2 \cos q$, i.e., $q=\cos^{-1}(1+i\omega/2)$ and $q^*=\cos^{-1}(1-i\omega/2)$ from the tri-diagonal form of $(\Phi+i\omega \mathds{1})$ matrix we can write~\cite{hu_1996} 
		\begin{equation}
			\mathcal{G}_{l,m}(\omega)=[(\Phi+i\omega \mathds{1}) ^{-1}]_{l,m} = \frac{\sin lq ~\sin (N-m+1)q}{\sin q ~\sin(N+1)q}
		\end{equation}
		Using this we express the above sums as
		\begin{align}
			F_{l,m}(\omega) &= \sum_{k=1}^N \tilde{\mathcal{G}}_{l,k}(\omega)~ \tilde{\mathcal{G}}_{m,k}^{T}(-\omega) =  \frac{1}{|\sin q|^2 |\sin (N+1)q|^2} \left[\sin(N-l+1)q \sin(N-m+1)q^* \sum_{k=1}^{l} |\sin(kq)|^2 \right.\nonumber\\
			&\left.+ \sin lq \sin(N-m+1)q^* \sum_{k=l+1}^{m} \sin(N-k+1)q \sin(kq^*) +\sin(lq) \sin(mq^*) \sum_{k=m+1}^{N} |\sin(N-k+1)q|^2 \right]
		\end{align}
		After performing the summations we get
		\begin{align}
			F_{l,m}(\omega)&=
			\frac{i}{8} \frac{{\rm Im}\bigg(\Big[\cos (N q^*)-\cos \big((N+2) q^*\big)\Big] \Big[\cos \big((N+1-l-m)q\big)-\cos \big((N+1+l-m)q\big)\Big]\bigg)}
			{\sin (q) \sin (q^*) \sin \big((N+1) q\big) \sin \big((N+1) q^*\big) \sin \left(\frac{q-q^*}{2}\right) \sin \left(\frac{q+q^*}{2}\right)}\nonumber\\
		\end{align}
		{
			This can be further simplified to
			\begin{align}
				F_{l,m}(\omega) &= \frac{i}{4 (\sin \left(\frac{q-q^*}{2}\right) \sin (\frac{q+q^*}{2}))} \text{Im}\left[\frac{\cos \big((N+1-l-m)q\big)-\cos \big((N+1+l-m)q\big)}
				{\sin (q) \sin \big[(N+1) q\big)\big] }\right]\,,\nonumber\\
				&=-\frac{1}{\omega} \text{Im}\left[\frac{\sin (l q) \sin (N+1-m) q}
				{\sin (q) \sin \big((N+1) q\big)\big)}\right]\, ,
			\end{align}
			which is given in Eq.~\eqref{main_ggomega_ij} of the main text.}
		For the two-time correlation for the active particles, as we do not know the exact steady-state distribution we can use the only method of solving the equations of motion directly. With this we get from Eq.~\eqref{twotime_xcrl_matrixform}:
		\begin{align}
			C_{l,m}^x(t)&=\frac{2v_0^2 \alpha}{\pi} \int_{-\infty}^{\infty}  \left(\sum_s \frac{\phi_s(l)~ \phi_s(m)}{(\lambda_s^2+\omega^2)(4\alpha^2+\omega^2)}\right) e^{i\omega t}~d\omega \nonumber \\
			&=2v_0^2 \alpha \sum_s \frac{\phi_s(l) \phi_s(m)}{4\alpha^2 -\lambda_s^2}\left[\frac{e^{-\lambda_s t}}{\lambda_s}-\frac{e^{-2\alpha t}}{2\alpha}\right]
		\end{align}
		And the static correlation for the active case will be :
		\begin{equation}\label{app_eigenbasis_Sijx}
			S_{l,m}^x = 2v_0^2 \alpha \sum_s \frac{\phi_s(l) \phi_s(m)}{2\alpha \lambda_s (2\alpha+ \lambda_s)}\,,
		\end{equation}
		which in the large $\alpha$ limit reduces to Eq.~\eqref{eq_app_static_crl_phiinv} or \eqref{app_sij_eq_defn}.

\section{Calculation of residues for $\mathcal{B}(\omega)$ as in Eq.~\eqref{main_Bomega_trace_bulk}}\label{append_2}		
We recall that $\tilde{\mathcal{G}}(\omega)=(i\omega {\mathds{I}+\Phi)}^{-1} = \mathcal{A}^{-1}$. 
The eigenvalues $\lambda_q$ of the ${\Phi}$ matrix is given by
\begin{equation}
\lambda_{q}=2k(1-\cos q)\,, ~~~~q=\frac{s \pi}{N+1}, ~~s=1,2,\dots N\,.
\end{equation}
In terms of the eigenvalues we write Eq.~\eqref{main_Bomega_trace_bulk} as:
\begin{eqnarray}
\mathcal{B}(\omega)&=&\frac{1}{N}\sum_{l,k} \tilde{\mathcal{G}}_{l,k}(\omega) \tilde{\mathcal{G}}_{k,l}(-\omega) \nonumber\\
&=&\frac{1}{N} \sum_q \frac{1}{\lambda_{q}^2+\omega^2}\nonumber \\
&=&\frac{1}{2\pi}\int_{0}^{2\pi} \frac{dq}{4k^2(1-\cos q)^2+\omega^2}\,.
\end{eqnarray}

For this closed integral of $\mathcal{B}(\omega)$ we assume an unit circle with $z=e^{iq}$, then for the limits of $z$ it will be a closed integral of the form,
\begin{eqnarray}
\mathcal{B}(\omega)&=&\frac{1}{2\pi}\oint \frac{-iz dz}{k^2(z-1)^4+\omega^2z^2}\\
\mathcal{B}(\omega')&=& \frac{-i}{2\pi k^2} \oint \frac{z dz}{(z-1)^4+\omega^{'2}z^2}\,,
\end{eqnarray}
where $\omega'={\omega}/{k}$.
\par 
It has four poles corresponding to the roots of  
\begin{equation}
(z-1)^4+\omega'{^2}z^2=0. 
\end{equation}
Among them, one pair of roots is a complex conjugate of the other pair. They are given by  
\begin{equation}
z_1=\frac{(2+i\omega')-\sqrt{(2+i\omega')^2-4}}{2}, 
~~z_1^*=\frac{(2-i\omega')-\sqrt{(2-i\omega')^2-4}}{2}\,,
\end{equation}
and,
\begin{equation}
z_2=\frac{(2+i\omega')+\sqrt{(2+i\omega')^2-4}}{2}, 
~~z_2^*=\frac{(2-i\omega')+\sqrt{(2-i\omega')^2-4}}{2}\,.
\end{equation}
		%
The absolute values of the roots $z_1$ and $z_1^*$ decrease from $1$ towards $0$ as $\omega'$ increases from $0$ to $\pm \infty$. For the other pair, the trend is the opposite. Their amplitude  increases from $1$ as $\omega'$ changes from $0$ to $\pm \infty$. Thus, only $z_1$, $z_1^*$ remain within the unit circle contour.		
\par
Thus,
\begin{equation}
I (\omega') = \frac{-i}{2\pi k^2} \times 2\pi i \Big[{\rm{Res}}_{z=z_1}f(z,\omega')+ {\rm{Res}}_{z=z_1^*}f(z,\omega')\Big] \,,\\
\end{equation}
leading to
\begin{equation}
\mathcal{B}(\omega')=\frac{1}{k^2}\Big[\frac{-1}{2i\omega'\sqrt{4i\omega'-\omega^{'2}}}+\frac{-1}{-2i\omega'\sqrt{-4i\omega'-\omega^{'2}}}\Big]\,.
\end{equation}
Replacing $\omega=\omega'k$ we obtain:
\begin{equation}\label{iomega_final}
\mathcal{B}(\omega)=\frac{1}{2}\Big[\frac{-1}{i\omega\sqrt{4ik\omega-\omega^2}}+\frac{-1}{-i\omega\sqrt{-4ik\omega -\omega^2}}\Big]
\end{equation}
This is a real quantity and can be written in a more compact form as
\begin{equation}\label{app_Bomega_finalform}
\mathcal{B}(\omega)= 2 ~\text{Re}\left[\frac{1}{2\omega \sqrt{\omega^2-4ik\omega}}\right]\,,
\end{equation}
which is used in Eq.~\eqref{main_Bomega_finalform}.

		\section{Steady-state equal time correlation matrix}\label{append_ss_t0_correlmatrix}
\begin{figure}[htb]
			\centering
			\hskip -0.2 cm
			\vskip -0.01 cm
			\includegraphics*[width=11.80cm,height=5.50cm]{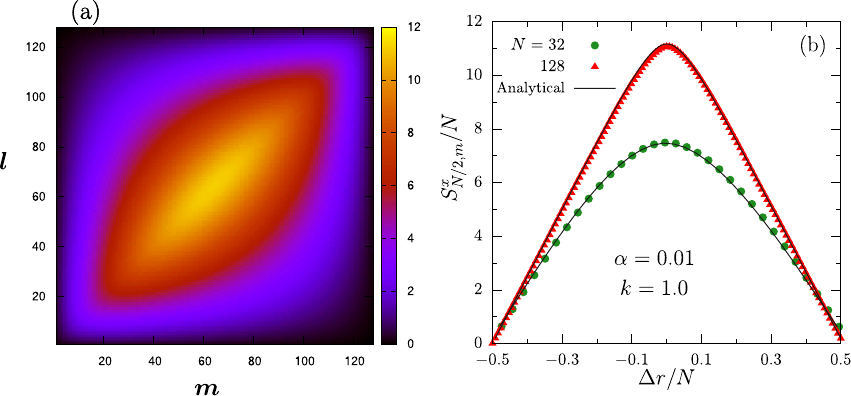}
			\caption{\label{appfig_heatmap_t0_crl} (a) Heat-map plot of the scaled  $S_{l,m}^x/N$ matrix for $\alpha=0.01$ and $N=128$ as presented in Eq.~\eqref{main_sscrl_t0_eq}. (b) Plot shows the $S_{N/2,m}^x/N$ vs $\Delta r/N=(m-N/2)/N$ for $N=32$ and $128$ and the solid lines are the analytical results of Eq.~\eqref{main_sscrl_t0_eq} with representation $R_2$.}
		\end{figure}

		By numerically evaluating the integral in the second representation of Eq.~\eqref{main_sscrl_t0_eq} with $F_{l,m}(\omega)$, in Fig.~\ref{appfig_heatmap_t0_crl}(a) we plot the heat map of the correlation matrix $S_{l,m}^x$ for $\alpha=0.01$ for a chain length $N=128$.
		As mentioned, we have calculated the correlation function between two points $l$ and $m$ by fixing $l=N/2$ and varying $m$ from $1$ to $N$. In Fig.~\ref{appfig_heatmap_t0_crl}(b) we show simulation results for $S_{l,m}^x$ for two chain lengths $N=32$ and $128$. In this, the solid lines are the plots of  Eq.~\eqref{main_sscrl_t0_eq}.

		\section{Time dependence of two-time correlations for displacement}
		\label{app_Cx}
\begin{figure}[htb]
			\centering
			\hskip 0.5 cm
			\includegraphics*[width=9cm]{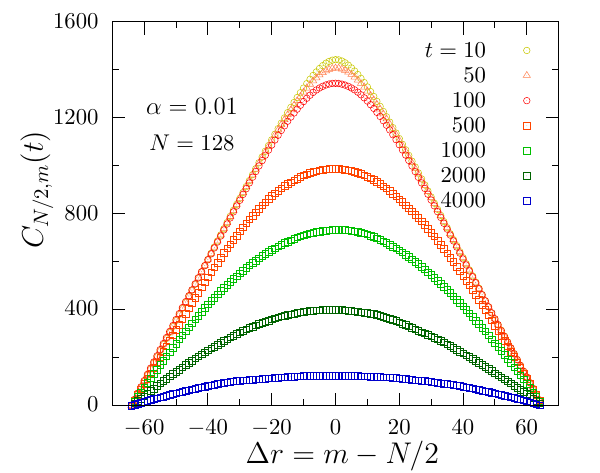}
			\caption{\label{xx_t1t2_alpha0.01} (a) Plot of two-time correlations $C_{l,m}^x(t)$ with $l=N/2$ and $m=N/2+\Delta r$ as a function of the separation $\Delta r$ between two particles for different times $t$, using $\alpha=0.01$ for $N=128$.}
		\end{figure}

		The two-time correlation of the displacement $x$, i.e., $C_{l,m}^x(t)$ was presented in Eq.~\eqref{eq_two_point_correl} or \eqref{crl_t_rep1} of the main text.
		%
		The fixed boundary condition ensures that $C^x_{l=N/2,m}=0$ for $m=1, N$. In Fig.~\ref{xx_t1t2_alpha0.01} we plot the time evolution of $C_{l=N/2,m}^x(t)$ vs $\Delta r=m-N/2$ for $\alpha=0.01$, which shows the spread of the correlation with increasing time.

		\section{Elements of the Green's function $\mathcal{\Tilde{G}}(\omega)$ matrix}
		\label{gmatrix_ij}
		Here we discuss a few properties of the $\mathcal{\Tilde{G}}$ and $\mathcal{A}=(\Phi+i\omega \mathds{1})$ matrix. 
		Now explicitly the $\mathcal{\Tilde{G}}_{l,m}(\omega)=\mathcal{A}^{-1}$ is
		\begin{equation}
			\mathcal{\Tilde{G}}_{l,m}(\omega) =\frac{\text{cofactor}[\mathcal{A}_{N}]}{\text{det}(\mathcal{A}_{N})}
		\end{equation}
		Recall that $\Phi$ has the form 
		\begin{equation}
			\Phi_{l,m}=k(2\delta_{l,m}-\delta_{l,m-1}-\delta_{l,m+1})\,.
		\end{equation}
		The elements of the $\mathcal{A}$ matrix is written as
		\begin{equation}
			\mathcal{A}_{l,m}=(a\delta_{l,m}-\delta_{l,m-1}-\delta_{l,m+1})\,,    
		\end{equation}
		where $a=(-2+i\omega/k)$. Now we consider $\mathcal{D}_{l,m}$ is the determinant of the sub-matrix of $\mathcal{A}$  starting with $l$-th row and column and ending with $m$-th row and column.  For this, it is simply obtained as \cite{dhar_2011}, 
		\begin{equation}
			\mathcal{D}_{l,m} = \frac{\sin (m-l+2)q}{\sin q}\,,
		\end{equation}
		where we have assumed $2+i\omega/k=2\cos q$. Now the $l,m$-th element of the $(\Phi+i\omega \mathds{1})^{-1}$, i.e., $\Tilde{\mathcal{G}}(\omega)$, we write
		\begin{equation}
			\mathcal{\Tilde{G}}_{l,m}(\omega) =\frac{\text{cofactor}[\mathcal{A}_{l,m}]}{\text{det}(\mathcal{A}_{N})}\,.
		\end{equation}
		Now to find the co-factor[$\mathcal{A}_{l,m}$] when we strike out the $l$-th row and the $m$-th column of the $\mathcal{A}$ matrix then the corresponding determinant can be evaluated by using the determinants of the three sub-block matrices of orders $(l-1)$, $(m-l)$ and $(N-m)$. We denote these matrices by $R$. Whether $R$ with orders $(l-1)$ and $(N-m)$ have the same form of $\mathcal{A}$ matrix, the $R$ matrix of the order $(m-l)$ has an upper diagonal form with all the diagonal elements as $1$. Writing in terms of the determinant $\mathcal{D}$ we get the $(l,m)$-th element of the $\Tilde{\mathcal{G}}$ as
		\begin{eqnarray}
			\mathcal{\Tilde{G}}_{l,m}(\omega)&=&\frac{\mathcal{D}_{1,l-1} \mathcal{D}_{N-m,N}}{\mathcal{D}_{1,N}} \nonumber \\
			&=& \frac{\sin(lq)~ \sin(N-m+1)q}{\sin q~\sin(N+1)q}
		\end{eqnarray}
		And its hermitian conjugate:
		\begin{align}
			\mathcal{\Tilde{G}}^T_{l,m}(-\omega) = \frac{\sin(lq^*) ~\sin(N-m+1)q^*}{\sin q^*~\sin(N+1)q^*}
		\end{align}


\begin{thebibliography}{68}%
			\makeatletter
			\providecommand \@ifxundefined [1]{%
				\@ifx{#1\undefined}
			}%
			\providecommand \@ifnum [1]{%
				\ifnum #1\expandafter \@firstoftwo
				\else \expandafter \@secondoftwo
				\fi
			}%
			\providecommand \@ifx [1]{%
				\ifx #1\expandafter \@firstoftwo
				\else \expandafter \@secondoftwo
				\fi
			}%
			\providecommand \natexlab [1]{#1}%
			\providecommand \enquote  [1]{``#1''}%
			\providecommand \bibnamefont  [1]{#1}%
			\providecommand \bibfnamefont [1]{#1}%
			\providecommand \citenamefont [1]{#1}%
			\providecommand \href@noop [0]{\@secondoftwo}%
			\providecommand \href [0]{\begingroup \@sanitize@url \@href}%
			\providecommand \@href[1]{\@@startlink{#1}\@@href}%
			\providecommand \@@href[1]{\endgroup#1\@@endlink}%
			\providecommand \@sanitize@url [0]{\catcode `\\12\catcode `\$12\catcode
				`\&12\catcode `\#12\catcode `\^12\catcode `\_12\catcode `\%12\relax}%
			\providecommand \@@startlink[1]{}%
			\providecommand \@@endlink[0]{}%
			\providecommand \url  [0]{\begingroup\@sanitize@url \@url }%
			\providecommand \@url [1]{\endgroup\@href {#1}{\urlprefix }}%
			\providecommand \urlprefix  [0]{URL }%
			\providecommand \Eprint [0]{\href }%
			\providecommand \doibase [0]{https://doi.org/}%
			\providecommand \selectlanguage [0]{\@gobble}%
			\providecommand \bibinfo  [0]{\@secondoftwo}%
			\providecommand \bibfield  [0]{\@secondoftwo}%
			\providecommand \translation [1]{[#1]}%
			\providecommand \BibitemOpen [0]{}%
			\providecommand \bibitemStop [0]{}%
			\providecommand \bibitemNoStop [0]{.\EOS\space}%
			\providecommand \EOS [0]{\spacefactor3000\relax}%
			\providecommand \BibitemShut  [1]{\csname bibitem#1\endcsname}%
			\let\auto@bib@innerbib\@empty
			\bibitem [{\citenamefont {Chat{\'e}}\ \emph {et~al.}(2008)\citenamefont
				{Chat{\'e}}, \citenamefont {Ginelli}, \citenamefont {Gr{\'e}goire},
				\citenamefont {Peruani},\ and\ \citenamefont {Raynaud}}]{chate_2008}%
			\BibitemOpen
			\bibfield  {author} {\bibinfo {author} {\bibfnamefont {H.}~\bibnamefont
					{Chat{\'e}}}, \bibinfo {author} {\bibfnamefont {F.}~\bibnamefont {Ginelli}},
				\bibinfo {author} {\bibfnamefont {G.}~\bibnamefont {Gr{\'e}goire}}, \bibinfo
				{author} {\bibfnamefont {F.}~\bibnamefont {Peruani}},\ and\ \bibinfo {author}
				{\bibfnamefont {F.}~\bibnamefont {Raynaud}},\ }\bibfield  {title} {\bibinfo
				{title} {Modeling collective motion: {V}ariations on the {V}icsek model},\
			}\href@noop {} {\bibfield  {journal} {\bibinfo  {journal} {The Eur. Phys. J.
						B}\ }\textbf {\bibinfo {volume} {64}},\ \bibinfo {pages} {451} (\bibinfo
				{year} {2008})}\BibitemShut {NoStop}%
			\bibitem [{\citenamefont {Ramaswamy}(2010)}]{ramaswamy2010}%
			\BibitemOpen
			\bibfield  {author} {\bibinfo {author} {\bibfnamefont {S.}~\bibnamefont
					{Ramaswamy}},\ }\bibfield  {title} {\bibinfo {title} {The mechanics and
					statistics of active matter},\ }\href
			{https://doi.org/10.1146/annurev-conmatphys-070909-104101} {\bibfield
				{journal} {\bibinfo  {journal} {Ann. Rev. Cond. Mat. Phys.}\ }\textbf
				{\bibinfo {volume} {1}},\ \bibinfo {pages} {323} (\bibinfo {year}
				{2010})}\BibitemShut {NoStop}%
			\bibitem [{\citenamefont {Romanczuk}\ \emph {et~al.}(2012)\citenamefont
				{Romanczuk}, \citenamefont {B{\"a}r}, \citenamefont {Ebeling}, \citenamefont
				{Lindner},\ and\ \citenamefont {Schimansky-Geier}}]{romanczuk_2012}%
			\BibitemOpen
			\bibfield  {author} {\bibinfo {author} {\bibfnamefont {P.}~\bibnamefont
					{Romanczuk}}, \bibinfo {author} {\bibfnamefont {M.}~\bibnamefont {B{\"a}r}},
				\bibinfo {author} {\bibfnamefont {W.}~\bibnamefont {Ebeling}}, \bibinfo
				{author} {\bibfnamefont {B.}~\bibnamefont {Lindner}},\ and\ \bibinfo {author}
				{\bibfnamefont {L.}~\bibnamefont {Schimansky-Geier}},\ }\bibfield  {title}
			{\bibinfo {title} {Active {B}rownian particles: From individual to collective
					stochastic dynamics},\ }\href@noop {} {\bibfield  {journal} {\bibinfo
					{journal} {The Eur. Phys. J. Spec. Top.}\ }\textbf {\bibinfo {volume}
					{202}},\ \bibinfo {pages} {1} (\bibinfo {year} {2012})}\BibitemShut {NoStop}%
			\bibitem [{\citenamefont {Marchetti}\ \emph {et~al.}(2013)\citenamefont
				{Marchetti}, \citenamefont {Joanny}, \citenamefont {Ramaswamy}, \citenamefont
				{Liverpool}, \citenamefont {Prost}, \citenamefont {Rao},\ and\ \citenamefont
				{Simha}}]{marchetti_13}%
			\BibitemOpen
			\bibfield  {author} {\bibinfo {author} {\bibfnamefont {M.~C.}\ \bibnamefont
					{Marchetti}}, \bibinfo {author} {\bibfnamefont {J.-F.}\ \bibnamefont
					{Joanny}}, \bibinfo {author} {\bibfnamefont {S.}~\bibnamefont {Ramaswamy}},
				\bibinfo {author} {\bibfnamefont {T.~B.}\ \bibnamefont {Liverpool}}, \bibinfo
				{author} {\bibfnamefont {J.}~\bibnamefont {Prost}}, \bibinfo {author}
				{\bibfnamefont {M.}~\bibnamefont {Rao}},\ and\ \bibinfo {author}
				{\bibfnamefont {R.~A.}\ \bibnamefont {Simha}},\ }\bibfield  {title} {\bibinfo
				{title} {Hydrodynamics of soft active matter},\ }\href@noop {} {\bibfield
				{journal} {\bibinfo  {journal} {Rev. Mod. Phys.}\ }\textbf {\bibinfo {volume}
					{85}},\ \bibinfo {pages} {1143} (\bibinfo {year} {2013})}\BibitemShut
			{NoStop}%
			\bibitem [{\citenamefont {Bechinger}\ \emph {et~al.}(2016)\citenamefont
				{Bechinger}, \citenamefont {Di~Leonardo}, \citenamefont {L\"owen},
				\citenamefont {Reichhardt}, \citenamefont {Volpe},\ and\ \citenamefont
				{Volpe}}]{beching_rmp_16}%
			\BibitemOpen
			\bibfield  {author} {\bibinfo {author} {\bibfnamefont {C.}~\bibnamefont
					{Bechinger}}, \bibinfo {author} {\bibfnamefont {R.}~\bibnamefont
					{Di~Leonardo}}, \bibinfo {author} {\bibfnamefont {H.}~\bibnamefont
					{L\"owen}}, \bibinfo {author} {\bibfnamefont {C.}~\bibnamefont {Reichhardt}},
				\bibinfo {author} {\bibfnamefont {G.}~\bibnamefont {Volpe}},\ and\ \bibinfo
				{author} {\bibfnamefont {G.}~\bibnamefont {Volpe}},\ }\bibfield  {title}
			{\bibinfo {title} {Active particles in complex and crowded environments},\
			}\href {https://doi.org/10.1103/RevModPhys.88.045006} {\bibfield  {journal}
				{\bibinfo  {journal} {Rev. Mod. Phys.}\ }\textbf {\bibinfo {volume} {88}},\
				\bibinfo {pages} {045006} (\bibinfo {year} {2016})}\BibitemShut {NoStop}%
			\bibitem [{\citenamefont {Loi}\ \emph {et~al.}(2008)\citenamefont {Loi},
				\citenamefont {Mossa},\ and\ \citenamefont {Cugliandolo}}]{loi_08}%
			\BibitemOpen
			\bibfield  {author} {\bibinfo {author} {\bibfnamefont {D.}~\bibnamefont
					{Loi}}, \bibinfo {author} {\bibfnamefont {S.}~\bibnamefont {Mossa}},\ and\
				\bibinfo {author} {\bibfnamefont {L.~F.}\ \bibnamefont {Cugliandolo}},\
			}\bibfield  {title} {\bibinfo {title} {Effective temperature of active
					matter},\ }\href {https://doi.org/10.1103/PhysRevE.77.051111} {\bibfield
				{journal} {\bibinfo  {journal} {Phys. Rev. E}\ }\textbf {\bibinfo {volume}
					{77}},\ \bibinfo {pages} {051111} (\bibinfo {year} {2008})}\BibitemShut
			{NoStop}%
			\bibitem [{\citenamefont {Fodor}\ \emph {et~al.}(2016)\citenamefont {Fodor},
				\citenamefont {Nardini}, \citenamefont {Cates}, \citenamefont {Tailleur},
				\citenamefont {Visco},\ and\ \citenamefont {van Wijland}}]{fodor_16}%
			\BibitemOpen
			\bibfield  {author} {\bibinfo {author} {\bibfnamefont {E.}~\bibnamefont
					{Fodor}}, \bibinfo {author} {\bibfnamefont {C.}~\bibnamefont {Nardini}},
				\bibinfo {author} {\bibfnamefont {M.~E.}\ \bibnamefont {Cates}}, \bibinfo
				{author} {\bibfnamefont {J.}~\bibnamefont {Tailleur}}, \bibinfo {author}
				{\bibfnamefont {P.}~\bibnamefont {Visco}},\ and\ \bibinfo {author}
				{\bibfnamefont {F.}~\bibnamefont {van Wijland}},\ }\bibfield  {title}
			{\bibinfo {title} {How far from equilibrium is active matter?},\ }\href
			{https://doi.org/10.1103/PhysRevLett.117.038103} {\bibfield  {journal}
				{\bibinfo  {journal} {Phys. Rev. Lett.}\ }\textbf {\bibinfo {volume} {117}},\
				\bibinfo {pages} {038103} (\bibinfo {year} {2016})}\BibitemShut {NoStop}%
			\bibitem [{\citenamefont {Astumian}\ and\ \citenamefont
				{H{\"{a}}nggi}(2002)}]{Astumian2002}%
			\BibitemOpen
			\bibfield  {author} {\bibinfo {author} {\bibfnamefont {R.~D.}\ \bibnamefont
					{Astumian}}\ and\ \bibinfo {author} {\bibfnamefont {P.}~\bibnamefont
					{H{\"{a}}nggi}},\ }\bibfield  {title} {\bibinfo {title} {{Brownian motors}},\
			}\href {https://doi.org/10.1063/1.1535005} {\bibfield  {journal} {\bibinfo
					{journal} {Physics Today}\ }\textbf {\bibinfo {volume} {55}},\ \bibinfo
				{pages} {33} (\bibinfo {year} {2002})}\BibitemShut {NoStop}%
			\bibitem [{\citenamefont {Berg}\ and\ \citenamefont {Brown}(1972)}]{Berg1972}%
			\BibitemOpen
			\bibfield  {author} {\bibinfo {author} {\bibfnamefont {H.~C.}\ \bibnamefont
					{Berg}}\ and\ \bibinfo {author} {\bibfnamefont {D.~A.}\ \bibnamefont
					{Brown}},\ }\bibfield  {title} {\bibinfo {title} {{Chemotaxis in Escherichia
						coli analysed by three-dimensional tracking}},\ }\href
			{https://doi.org/10.1038/239500a0} {\bibfield  {journal} {\bibinfo  {journal}
					{Nature}\ }\textbf {\bibinfo {volume} {239}},\ \bibinfo {pages} {500}
				(\bibinfo {year} {1972})}\BibitemShut {NoStop}%
			\bibitem [{\citenamefont {Niwa}(1994)}]{Niwa1994}%
			\BibitemOpen
			\bibfield  {author} {\bibinfo {author} {\bibfnamefont {H.~S.}\ \bibnamefont
					{Niwa}},\ }\bibfield  {title} {\bibinfo {title} {{Self-organizing dynamic
						model of fish schooling}},\ }\href {https://doi.org/10.1006/jtbi.1994.1218}
			{\bibfield  {journal} {\bibinfo  {journal} {J. Theor. Biol.}\ }\textbf
				{\bibinfo {volume} {171}},\ \bibinfo {pages} {123} (\bibinfo {year}
				{1994})}\BibitemShut {NoStop}%
			\bibitem [{\citenamefont {Ginelli}\ \emph {et~al.}(2015)\citenamefont
				{Ginelli}, \citenamefont {Peruani}, \citenamefont {Pillot}, \citenamefont
				{Chat{\'{e}}}, \citenamefont {Theraulaz},\ and\ \citenamefont
				{Bon}}]{Ginelli2015}%
			\BibitemOpen
			\bibfield  {author} {\bibinfo {author} {\bibfnamefont {F.}~\bibnamefont
					{Ginelli}}, \bibinfo {author} {\bibfnamefont {F.}~\bibnamefont {Peruani}},
				\bibinfo {author} {\bibfnamefont {M.-H.}\ \bibnamefont {Pillot}}, \bibinfo
				{author} {\bibfnamefont {H.}~\bibnamefont {Chat{\'{e}}}}, \bibinfo {author}
				{\bibfnamefont {G.}~\bibnamefont {Theraulaz}},\ and\ \bibinfo {author}
				{\bibfnamefont {R.}~\bibnamefont {Bon}},\ }\bibfield  {title} {\bibinfo
				{title} {{Intermittent collective dynamics emerge from conflicting
						imperatives in sheep herds}},\ }\href
			{https://doi.org/10.1073/pnas.1503749112} {\bibfield  {journal} {\bibinfo
					{journal} {Proc. Natl. Acad. Sci.}\ }\textbf {\bibinfo {volume} {112}},\
				\bibinfo {pages} {12729} (\bibinfo {year} {2015})}\BibitemShut {NoStop}%
			\bibitem [{\citenamefont {Devereux}\ \emph {et~al.}(2021)\citenamefont
				{Devereux}, \citenamefont {Twomey}, \citenamefont {Turner},\ and\
				\citenamefont {Thutupalli}}]{Devereux2021}%
			\BibitemOpen
			\bibfield  {author} {\bibinfo {author} {\bibfnamefont {H.~L.}\ \bibnamefont
					{Devereux}}, \bibinfo {author} {\bibfnamefont {C.~R.}\ \bibnamefont
					{Twomey}}, \bibinfo {author} {\bibfnamefont {M.~S.}\ \bibnamefont {Turner}},\
				and\ \bibinfo {author} {\bibfnamefont {S.}~\bibnamefont {Thutupalli}},\
			}\bibfield  {title} {\bibinfo {title} {{Whirligig beetles as corralled active
						Brownian particles}},\ }\bibfield  {journal} {\bibinfo  {journal} {J. R. Soc.
					Interface}\ }\textbf {\bibinfo {volume} {18}},\ \href
			{https://doi.org/10.1098/rsif.2021.0114} {10.1098/rsif.2021.0114} (\bibinfo
			{year} {2021})\BibitemShut {NoStop}%
			\bibitem [{\citenamefont {Illien}\ \emph {et~al.}(2017)\citenamefont {Illien},
				\citenamefont {Golestanian},\ and\ \citenamefont {Sen}}]{Illien2017}%
			\BibitemOpen
			\bibfield  {author} {\bibinfo {author} {\bibfnamefont {P.}~\bibnamefont
					{Illien}}, \bibinfo {author} {\bibfnamefont {R.}~\bibnamefont
					{Golestanian}},\ and\ \bibinfo {author} {\bibfnamefont {A.}~\bibnamefont
					{Sen}},\ }\bibfield  {title} {\bibinfo {title} {{`Fuelled' motion: Phoretic
						motility and collective behaviour of active colloids}},\ }\href
			{https://doi.org/10.1039/c7cs00087a} {\bibfield  {journal} {\bibinfo
					{journal} {Chem. Soc. Rev.}\ }\textbf {\bibinfo {volume} {46}},\ \bibinfo
				{pages} {5508} (\bibinfo {year} {2017})}\BibitemShut {NoStop}%
			\bibitem [{\citenamefont {Paxton}\ \emph {et~al.}(2004)\citenamefont {Paxton},
				\citenamefont {Kistler}, \citenamefont {Olmeda}, \citenamefont {Sen},
				\citenamefont {{St. Angelo}}, \citenamefont {Cao}, \citenamefont {Mallouk},
				\citenamefont {Lammert},\ and\ \citenamefont {Crespi}}]{Paxton2004}%
			\BibitemOpen
			\bibfield  {author} {\bibinfo {author} {\bibfnamefont {W.~F.}\ \bibnamefont
					{Paxton}}, \bibinfo {author} {\bibfnamefont {K.~C.}\ \bibnamefont {Kistler}},
				\bibinfo {author} {\bibfnamefont {C.~C.}\ \bibnamefont {Olmeda}}, \bibinfo
				{author} {\bibfnamefont {A.}~\bibnamefont {Sen}}, \bibinfo {author}
				{\bibfnamefont {S.~K.}\ \bibnamefont {{St. Angelo}}}, \bibinfo {author}
				{\bibfnamefont {Y.}~\bibnamefont {Cao}}, \bibinfo {author} {\bibfnamefont
					{T.~E.}\ \bibnamefont {Mallouk}}, \bibinfo {author} {\bibfnamefont {P.~E.}\
					\bibnamefont {Lammert}},\ and\ \bibinfo {author} {\bibfnamefont {V.~H.}\
					\bibnamefont {Crespi}},\ }\bibfield  {title} {\bibinfo {title} {{Catalytic
						nanomotors: Autonomous movement of striped nanorods}},\ }\href
			{https://doi.org/10.1021/ja047697z} {\bibfield  {journal} {\bibinfo
					{journal} {J. Am. Chem. Soc.}\ }\textbf {\bibinfo {volume} {126}},\ \bibinfo
				{pages} {13424} (\bibinfo {year} {2004})}\BibitemShut {NoStop}%
			\bibitem [{\citenamefont {Bricard}\ \emph {et~al.}(2013)\citenamefont
				{Bricard}, \citenamefont {Caussin}, \citenamefont {Desreumaux}, \citenamefont
				{Dauchot},\ and\ \citenamefont {Bartolo}}]{Bricard2013}%
			\BibitemOpen
			\bibfield  {author} {\bibinfo {author} {\bibfnamefont {A.}~\bibnamefont
					{Bricard}}, \bibinfo {author} {\bibfnamefont {J.~B.}\ \bibnamefont
					{Caussin}}, \bibinfo {author} {\bibfnamefont {N.}~\bibnamefont {Desreumaux}},
				\bibinfo {author} {\bibfnamefont {O.}~\bibnamefont {Dauchot}},\ and\ \bibinfo
				{author} {\bibfnamefont {D.}~\bibnamefont {Bartolo}},\ }\bibfield  {title}
			{\bibinfo {title} {{Emergence of macroscopic directed motion in populations
						of motile colloids}},\ }\href {https://doi.org/10.1038/nature12673}
			{\bibfield  {journal} {\bibinfo  {journal} {Nature}\ }\textbf {\bibinfo
					{volume} {503}},\ \bibinfo {pages} {95} (\bibinfo {year} {2013})}\BibitemShut
			{NoStop}%
			\bibitem [{\citenamefont {Bricard}\ \emph {et~al.}(2015)\citenamefont
				{Bricard}, \citenamefont {Caussin}, \citenamefont {Das}, \citenamefont
				{Savoie}, \citenamefont {Chikkadi}, \citenamefont {Shitara}, \citenamefont
				{Chepizhko}, \citenamefont {Peruani}, \citenamefont {Saintillan},\ and\
				\citenamefont {Bartolo}}]{Bricard2015}%
			\BibitemOpen
			\bibfield  {author} {\bibinfo {author} {\bibfnamefont {A.}~\bibnamefont
					{Bricard}}, \bibinfo {author} {\bibfnamefont {J.-B.}\ \bibnamefont
					{Caussin}}, \bibinfo {author} {\bibfnamefont {D.}~\bibnamefont {Das}},
				\bibinfo {author} {\bibfnamefont {C.}~\bibnamefont {Savoie}}, \bibinfo
				{author} {\bibfnamefont {V.}~\bibnamefont {Chikkadi}}, \bibinfo {author}
				{\bibfnamefont {K.}~\bibnamefont {Shitara}}, \bibinfo {author} {\bibfnamefont
					{O.}~\bibnamefont {Chepizhko}}, \bibinfo {author} {\bibfnamefont
					{F.}~\bibnamefont {Peruani}}, \bibinfo {author} {\bibfnamefont
					{D.}~\bibnamefont {Saintillan}},\ and\ \bibinfo {author} {\bibfnamefont
					{D.}~\bibnamefont {Bartolo}},\ }\bibfield  {title} {\bibinfo {title}
				{{Emergent vortices in populations of colloidal rollers}},\ }\href
			{https://doi.org/10.1038/ncomms8470} {\bibfield  {journal} {\bibinfo
					{journal} {Nat. Commun.}\ }\textbf {\bibinfo {volume} {6}},\ \bibinfo {pages}
				{7470} (\bibinfo {year} {2015})}\BibitemShut {NoStop}%
			\bibitem [{\citenamefont {Dauchot}\ and\ \citenamefont
				{D{\'{e}}mery}(2019)}]{Dauchot2019}%
			\BibitemOpen
			\bibfield  {author} {\bibinfo {author} {\bibfnamefont {O.}~\bibnamefont
					{Dauchot}}\ and\ \bibinfo {author} {\bibfnamefont {V.}~\bibnamefont
					{D{\'{e}}mery}},\ }\bibfield  {title} {\bibinfo {title} {{Dynamics of a
						Self-Propelled Particle in a Harmonic Trap}},\ }\href
			{https://doi.org/10.1103/PhysRevLett.122.068002} {\bibfield  {journal}
				{\bibinfo  {journal} {Phys. Rev. Lett.}\ }\textbf {\bibinfo {volume} {122}},\
				\bibinfo {pages} {068002} (\bibinfo {year} {2019})}\BibitemShut {NoStop}%
			\bibitem [{\citenamefont {Scholz}\ \emph {et~al.}(2018)\citenamefont {Scholz},
				\citenamefont {Engel},\ and\ \citenamefont {P{\"{o}}schel}}]{Scholz2018}%
			\BibitemOpen
			\bibfield  {author} {\bibinfo {author} {\bibfnamefont {C.}~\bibnamefont
					{Scholz}}, \bibinfo {author} {\bibfnamefont {M.}~\bibnamefont {Engel}},\ and\
				\bibinfo {author} {\bibfnamefont {T.}~\bibnamefont {P{\"{o}}schel}},\
			}\bibfield  {title} {\bibinfo {title} {{Rotating robots move collectively and
						self-organize}},\ }\href {https://doi.org/10.1038/s41467-018-03154-7}
			{\bibfield  {journal} {\bibinfo  {journal} {Nat. Commun.}\ }\textbf {\bibinfo
					{volume} {9}},\ \bibinfo {pages} {931} (\bibinfo {year} {2018})}\BibitemShut
			{NoStop}%
			\bibitem [{\citenamefont {Narayan}\ \emph {et~al.}(2007)\citenamefont
				{Narayan}, \citenamefont {Ramaswamy}, \citenamefont {Menon}, \citenamefont
				{Caspt},\ and\ \citenamefont {Caspt}}]{Narayan2007}%
			\BibitemOpen
			\bibfield  {author} {\bibinfo {author} {\bibfnamefont {V.}~\bibnamefont
					{Narayan}}, \bibinfo {author} {\bibfnamefont {S.}~\bibnamefont {Ramaswamy}},
				\bibinfo {author} {\bibfnamefont {N.}~\bibnamefont {Menon}}, \bibinfo
				{author} {\bibfnamefont {T.}~\bibnamefont {Caspt}},\ and\ \bibinfo {author}
				{\bibfnamefont {T.}~\bibnamefont {Caspt}},\ }\bibfield  {title} {\bibinfo
				{title} {{Long-Lived Giant Number Fluctuations}},\ }\href
			{http://www.ncbi.nlm.nih.gov/pubmed/17615353} {\bibfield  {journal} {\bibinfo
					{journal} {Science}\ }\textbf {\bibinfo {volume} {317}},\ \bibinfo {pages}
				{105} (\bibinfo {year} {2007})}\BibitemShut {NoStop}%
			\bibitem [{\citenamefont {Kudrolli}\ \emph {et~al.}(2008)\citenamefont
				{Kudrolli}, \citenamefont {Lumay}, \citenamefont {Volfson},\ and\
				\citenamefont {Tsimring}}]{Kudrolli2008}%
			\BibitemOpen
			\bibfield  {author} {\bibinfo {author} {\bibfnamefont {A.}~\bibnamefont
					{Kudrolli}}, \bibinfo {author} {\bibfnamefont {G.}~\bibnamefont {Lumay}},
				\bibinfo {author} {\bibfnamefont {D.}~\bibnamefont {Volfson}},\ and\ \bibinfo
				{author} {\bibfnamefont {L.~S.}\ \bibnamefont {Tsimring}},\ }\bibfield
			{title} {\bibinfo {title} {{Swarming and Swirling in Self-Propelled Polar
						Granular Rods}},\ }\href {https://doi.org/10.1103/PhysRevLett.100.058001}
			{\bibfield  {journal} {\bibinfo  {journal} {Phys. Rev. Lett.}\ }\textbf
				{\bibinfo {volume} {100}},\ \bibinfo {pages} {058001} (\bibinfo {year}
				{2008})}\BibitemShut {NoStop}%
			\bibitem [{\citenamefont {Deseigne}\ \emph {et~al.}(2010)\citenamefont
				{Deseigne}, \citenamefont {Dauchot},\ and\ \citenamefont
				{Chat{\'{e}}}}]{Deseigne2010}%
			\BibitemOpen
			\bibfield  {author} {\bibinfo {author} {\bibfnamefont {J.}~\bibnamefont
					{Deseigne}}, \bibinfo {author} {\bibfnamefont {O.}~\bibnamefont {Dauchot}},\
				and\ \bibinfo {author} {\bibfnamefont {H.}~\bibnamefont {Chat{\'{e}}}},\
			}\bibfield  {title} {\bibinfo {title} {{Collective Motion of Vibrated Polar
						Disks}},\ }\href {https://doi.org/10.1103/PhysRevLett.105.098001} {\bibfield
				{journal} {\bibinfo  {journal} {Phys. Rev. Lett.}\ }\textbf {\bibinfo
					{volume} {105}},\ \bibinfo {pages} {098001} (\bibinfo {year}
				{2010})}\BibitemShut {NoStop}%
			\bibitem [{\citenamefont {Farhadi}\ \emph {et~al.}(2018)\citenamefont
				{Farhadi}, \citenamefont {Machaca}, \citenamefont {Aird}, \citenamefont
				{{Torres Maldonado}}, \citenamefont {Davis}, \citenamefont {Arratia},\ and\
				\citenamefont {Durian}}]{Farhadi2018}%
			\BibitemOpen
			\bibfield  {author} {\bibinfo {author} {\bibfnamefont {S.}~\bibnamefont
					{Farhadi}}, \bibinfo {author} {\bibfnamefont {S.}~\bibnamefont {Machaca}},
				\bibinfo {author} {\bibfnamefont {J.}~\bibnamefont {Aird}}, \bibinfo {author}
				{\bibfnamefont {B.~O.}\ \bibnamefont {{Torres Maldonado}}}, \bibinfo {author}
				{\bibfnamefont {S.}~\bibnamefont {Davis}}, \bibinfo {author} {\bibfnamefont
					{P.~E.}\ \bibnamefont {Arratia}},\ and\ \bibinfo {author} {\bibfnamefont
					{D.~J.}\ \bibnamefont {Durian}},\ }\bibfield  {title} {\bibinfo {title}
				{{Dynamics and thermodynamics of air-driven active spinners}},\ }\href
			{https://doi.org/10.1039/c8sm00403j} {\bibfield  {journal} {\bibinfo
					{journal} {Soft Matter}\ }\textbf {\bibinfo {volume} {14}},\ \bibinfo {pages}
				{5588} (\bibinfo {year} {2018})}\BibitemShut {NoStop}%
			\bibitem [{\citenamefont {Stenhammar}\ \emph {et~al.}(2013)\citenamefont
				{Stenhammar}, \citenamefont {Tiribocchi}, \citenamefont {Allen},
				\citenamefont {Marenduzzo},\ and\ \citenamefont {Cates}}]{stenhamm2013}%
			\BibitemOpen
			\bibfield  {author} {\bibinfo {author} {\bibfnamefont {J.}~\bibnamefont
					{Stenhammar}}, \bibinfo {author} {\bibfnamefont {A.}~\bibnamefont
					{Tiribocchi}}, \bibinfo {author} {\bibfnamefont {R.~J.}\ \bibnamefont
					{Allen}}, \bibinfo {author} {\bibfnamefont {D.}~\bibnamefont {Marenduzzo}},\
				and\ \bibinfo {author} {\bibfnamefont {M.~E.}\ \bibnamefont {Cates}},\
			}\bibfield  {title} {\bibinfo {title} {Continuum theory of phase separation
					kinetics for active {B}rownian particles},\ }\href
			{https://doi.org/10.1103/PhysRevLett.111.145702} {\bibfield  {journal}
				{\bibinfo  {journal} {Phys. Rev. Lett.}\ }\textbf {\bibinfo {volume} {111}},\
				\bibinfo {pages} {145702} (\bibinfo {year} {2013})}\BibitemShut {NoStop}%
			\bibitem [{\citenamefont {Cates}\ and\ \citenamefont
				{Tailleur}(2015)}]{cates_mips_15}%
			\BibitemOpen
			\bibfield  {author} {\bibinfo {author} {\bibfnamefont {M.~E.}\ \bibnamefont
					{Cates}}\ and\ \bibinfo {author} {\bibfnamefont {J.}~\bibnamefont
					{Tailleur}},\ }\bibfield  {title} {\bibinfo {title} {Motility-induced phase
					separation},\ }\href
			{https://doi.org/10.1146/annurev-conmatphys-031214-014710} {\bibfield
				{journal} {\bibinfo  {journal} {Ann. Rev. Cond. Mat. Phys.}\ }\textbf
				{\bibinfo {volume} {6}},\ \bibinfo {pages} {219} (\bibinfo {year}
				{2015})}\BibitemShut {NoStop}%
			\bibitem [{\citenamefont {Tailleur}\ and\ \citenamefont
				{Cates}(2008)}]{tailleur_rtp_08}%
			\BibitemOpen
			\bibfield  {author} {\bibinfo {author} {\bibfnamefont {J.}~\bibnamefont
					{Tailleur}}\ and\ \bibinfo {author} {\bibfnamefont {M.~E.}\ \bibnamefont
					{Cates}},\ }\bibfield  {title} {\bibinfo {title} {Statistical mechanics of
					interacting run-and-tumble bacteria},\ }\href
			{https://doi.org/10.1103/PhysRevLett.100.218103} {\bibfield  {journal}
				{\bibinfo  {journal} {Phys. Rev. Lett.}\ }\textbf {\bibinfo {volume} {100}},\
				\bibinfo {pages} {218103} (\bibinfo {year} {2008})}\BibitemShut {NoStop}%
			\bibitem [{\citenamefont {Basu}\ \emph {et~al.}(2018)\citenamefont {Basu},
				\citenamefont {Majumdar}, \citenamefont {Rosso},\ and\ \citenamefont
				{Schehr}}]{basu2018}%
			\BibitemOpen
			\bibfield  {author} {\bibinfo {author} {\bibfnamefont {U.}~\bibnamefont
					{Basu}}, \bibinfo {author} {\bibfnamefont {S.~N.}\ \bibnamefont {Majumdar}},
				\bibinfo {author} {\bibfnamefont {A.}~\bibnamefont {Rosso}},\ and\ \bibinfo
				{author} {\bibfnamefont {G.}~\bibnamefont {Schehr}},\ }\bibfield  {title}
			{\bibinfo {title} {Active {B}rownian motion in two dimensions},\ }\href
			{https://doi.org/10.1103/PhysRevE.98.062121} {\bibfield  {journal} {\bibinfo
					{journal} {Phys. Rev. E}\ }\textbf {\bibinfo {volume} {98}},\ \bibinfo
				{pages} {062121} (\bibinfo {year} {2018})}\BibitemShut {NoStop}%
			\bibitem [{\citenamefont {Solon}\ \emph {et~al.}(2015)\citenamefont {Solon},
				\citenamefont {Fily}, \citenamefont {Baskaran}, \citenamefont {Cates},
				\citenamefont {Kafri}, \citenamefont {Kardar},\ and\ \citenamefont
				{Tailleur}}]{solon2015}%
			\BibitemOpen
			\bibfield  {author} {\bibinfo {author} {\bibfnamefont {A.~P.}\ \bibnamefont
					{Solon}}, \bibinfo {author} {\bibfnamefont {Y.}~\bibnamefont {Fily}},
				\bibinfo {author} {\bibfnamefont {A.}~\bibnamefont {Baskaran}}, \bibinfo
				{author} {\bibfnamefont {M.~E.}\ \bibnamefont {Cates}}, \bibinfo {author}
				{\bibfnamefont {Y.}~\bibnamefont {Kafri}}, \bibinfo {author} {\bibfnamefont
					{M.}~\bibnamefont {Kardar}},\ and\ \bibinfo {author} {\bibfnamefont
					{J.}~\bibnamefont {Tailleur}},\ }\bibfield  {title} {\bibinfo {title}
				{Pressure is not a state function for generic active fluids},\ }\href@noop {}
			{\bibfield  {journal} {\bibinfo  {journal} {Nat. Phys.}\ }\textbf {\bibinfo
					{volume} {11}},\ \bibinfo {pages} {673} (\bibinfo {year} {2015})}\BibitemShut
			{NoStop}%
			\bibitem [{\citenamefont {Dhar}\ \emph {et~al.}(2019)\citenamefont {Dhar},
				\citenamefont {Kundu}, \citenamefont {Majumdar}, \citenamefont
				{Sabhapandit},\ and\ \citenamefont {Schehr}}]{Dhar2019}%
			\BibitemOpen
			\bibfield  {author} {\bibinfo {author} {\bibfnamefont {A.}~\bibnamefont
					{Dhar}}, \bibinfo {author} {\bibfnamefont {A.}~\bibnamefont {Kundu}},
				\bibinfo {author} {\bibfnamefont {S.~N.}\ \bibnamefont {Majumdar}}, \bibinfo
				{author} {\bibfnamefont {S.}~\bibnamefont {Sabhapandit}},\ and\ \bibinfo
				{author} {\bibfnamefont {G.}~\bibnamefont {Schehr}},\ }\bibfield  {title}
			{\bibinfo {title} {Run-and-tumble particle in one-dimensional confining
					potentials: Steady-state, relaxation, and first-passage properties},\ }\href
			{https://doi.org/10.1103/PhysRevE.99.032132} {\bibfield  {journal} {\bibinfo
					{journal} {Phys. Rev. E}\ }\textbf {\bibinfo {volume} {99}},\ \bibinfo
				{pages} {032132} (\bibinfo {year} {2019})}\BibitemShut {NoStop}%
			\bibitem [{\citenamefont {Patel}\ and\ \citenamefont
				{Chaudhuri}(2023)}]{Patel2023}%
			\BibitemOpen
			\bibfield  {author} {\bibinfo {author} {\bibfnamefont {M.}~\bibnamefont
					{Patel}}\ and\ \bibinfo {author} {\bibfnamefont {D.}~\bibnamefont
					{Chaudhuri}},\ }\bibfield  {title} {\bibinfo {title} {Exact moments and
					re-entrant transitions in the inertial dynamics of active {B}rownian
					particles},\ }\href {https://doi.org/10.1088/1367-2630/ad1538} {\bibfield
				{journal} {\bibinfo  {journal} {New J. Phys.}\ }\textbf {\bibinfo {volume}
					{25}},\ \bibinfo {pages} {123048} (\bibinfo {year} {2023})}\BibitemShut
			{NoStop}%
			\bibitem [{\citenamefont {Sevilla}\ and\ \citenamefont {{G{\'{o}}mez
						Nava}}(2014)}]{Sevilla2014}%
			\BibitemOpen
			\bibfield  {author} {\bibinfo {author} {\bibfnamefont {F.~J.}\ \bibnamefont
					{Sevilla}}\ and\ \bibinfo {author} {\bibfnamefont {L.~A.}\ \bibnamefont
					{{G{\'{o}}mez Nava}}},\ }\bibfield  {title} {\bibinfo {title} {{Theory of
						diffusion of active particles that move at constant speed in two
						dimensions}},\ }\href {https://doi.org/10.1103/PhysRevE.90.022130} {\bibfield
				{journal} {\bibinfo  {journal} {Phys. Rev. E}\ }\textbf {\bibinfo {volume}
					{90}},\ \bibinfo {pages} {22130} (\bibinfo {year} {2014})}\BibitemShut
			{NoStop}%
			\bibitem [{\citenamefont {Gro{\ss}mann}\ \emph {et~al.}(2016)\citenamefont
				{Gro{\ss}mann}, \citenamefont {Peruani},\ and\ \citenamefont
				{B{\"{a}}r}}]{Grossmann2016}%
			\BibitemOpen
			\bibfield  {author} {\bibinfo {author} {\bibfnamefont {R.}~\bibnamefont
					{Gro{\ss}mann}}, \bibinfo {author} {\bibfnamefont {F.}~\bibnamefont
					{Peruani}},\ and\ \bibinfo {author} {\bibfnamefont {M.}~\bibnamefont
					{B{\"{a}}r}},\ }\bibfield  {title} {\bibinfo {title} {{Diffusion properties
						of active particles with directional reversal}},\ }\href
			{https://doi.org/10.1088/1367-2630/18/4/043009} {\bibfield  {journal}
				{\bibinfo  {journal} {New J. Phys.}\ }\textbf {\bibinfo {volume} {18}},\
				\bibinfo {pages} {43009} (\bibinfo {year} {2016})}\BibitemShut {NoStop}%
			\bibitem [{\citenamefont {Kurzthaler}\ \emph {et~al.}(2018)\citenamefont
				{Kurzthaler}, \citenamefont {Devailly}, \citenamefont {Arlt}, \citenamefont
				{Franosch}, \citenamefont {Poon}, \citenamefont {Martinez},\ and\
				\citenamefont {Brown}}]{Kurzthaler2018a}%
			\BibitemOpen
			\bibfield  {author} {\bibinfo {author} {\bibfnamefont {C.}~\bibnamefont
					{Kurzthaler}}, \bibinfo {author} {\bibfnamefont {C.}~\bibnamefont
					{Devailly}}, \bibinfo {author} {\bibfnamefont {J.}~\bibnamefont {Arlt}},
				\bibinfo {author} {\bibfnamefont {T.}~\bibnamefont {Franosch}}, \bibinfo
				{author} {\bibfnamefont {W.~C.}\ \bibnamefont {Poon}}, \bibinfo {author}
				{\bibfnamefont {V.~A.}\ \bibnamefont {Martinez}},\ and\ \bibinfo {author}
				{\bibfnamefont {A.~T.}\ \bibnamefont {Brown}},\ }\bibfield  {title} {\bibinfo
				{title} {{Probing the Spatiotemporal Dynamics of Catalytic Janus Particles
						with Single-Particle Tracking and Differential Dynamic Microscopy}},\ }\href
			{https://doi.org/10.1103/PhysRevLett.121.078001} {\bibfield  {journal}
				{\bibinfo  {journal} {Phys. Rev. Lett.}\ }\textbf {\bibinfo {volume} {121}},\
				\bibinfo {pages} {1} (\bibinfo {year} {2018})}\BibitemShut {NoStop}%
			\bibitem [{\citenamefont {Malakar}\ \emph {et~al.}(2018)\citenamefont
				{Malakar}, \citenamefont {Jemseena}, \citenamefont {Kundu}, \citenamefont
				{Kumar}, \citenamefont {Sabhapandit}, \citenamefont {Majumdar}, \citenamefont
				{Redner},\ and\ \citenamefont {Dhar}}]{malakar_18}%
			\BibitemOpen
			\bibfield  {author} {\bibinfo {author} {\bibfnamefont {K.}~\bibnamefont
					{Malakar}}, \bibinfo {author} {\bibfnamefont {V.}~\bibnamefont {Jemseena}},
				\bibinfo {author} {\bibfnamefont {A.}~\bibnamefont {Kundu}}, \bibinfo
				{author} {\bibfnamefont {K.~V.}\ \bibnamefont {Kumar}}, \bibinfo {author}
				{\bibfnamefont {S.}~\bibnamefont {Sabhapandit}}, \bibinfo {author}
				{\bibfnamefont {S.~N.}\ \bibnamefont {Majumdar}}, \bibinfo {author}
				{\bibfnamefont {S.}~\bibnamefont {Redner}},\ and\ \bibinfo {author}
				{\bibfnamefont {A.}~\bibnamefont {Dhar}},\ }\bibfield  {title} {\bibinfo
				{title} {Steady state, relaxation and first-passage properties of a
					run-and-tumble particle in one-dimension},\ }\href@noop {} {\bibfield
				{journal} {\bibinfo  {journal} {J. Stat. Mech.: Theor. and Expt.}\ }\textbf
				{\bibinfo {volume} {2018}},\ \bibinfo {pages} {043215} (\bibinfo {year}
				{2018})}\BibitemShut {NoStop}%
			\bibitem [{\citenamefont {Basu}\ \emph {et~al.}(2019)\citenamefont {Basu},
				\citenamefont {Majumdar}, \citenamefont {Rosso},\ and\ \citenamefont
				{Schehr}}]{Basu2019}%
			\BibitemOpen
			\bibfield  {author} {\bibinfo {author} {\bibfnamefont {U.}~\bibnamefont
					{Basu}}, \bibinfo {author} {\bibfnamefont {S.~N.}\ \bibnamefont {Majumdar}},
				\bibinfo {author} {\bibfnamefont {A.}~\bibnamefont {Rosso}},\ and\ \bibinfo
				{author} {\bibfnamefont {G.}~\bibnamefont {Schehr}},\ }\bibfield  {title}
			{\bibinfo {title} {{Long-time position distribution of an active Brownian
						particle in two dimensions}},\ }\href
			{https://doi.org/10.1103/PhysRevE.100.062116} {\bibfield  {journal} {\bibinfo
					{journal} {Phys. Rev. E}\ }\textbf {\bibinfo {volume} {100}},\ \bibinfo
				{pages} {1} (\bibinfo {year} {2019})}\BibitemShut {NoStop}%
			\bibitem [{\citenamefont {Shee}\ \emph {et~al.}(2020)\citenamefont {Shee},
				\citenamefont {Dhar},\ and\ \citenamefont {Chaudhuri}}]{Shee2020}%
			\BibitemOpen
			\bibfield  {author} {\bibinfo {author} {\bibfnamefont {A.}~\bibnamefont
					{Shee}}, \bibinfo {author} {\bibfnamefont {A.}~\bibnamefont {Dhar}},\ and\
				\bibinfo {author} {\bibfnamefont {D.}~\bibnamefont {Chaudhuri}},\ }\bibfield
			{title} {\bibinfo {title} {{Active Brownian particles: {M}apping to
						equilibrium polymers and exact computation of moments}},\ }\href
			{https://doi.org/10.1039/D0SM00367K} {\bibfield  {journal} {\bibinfo
					{journal} {Soft Matter}\ }\textbf {\bibinfo {volume} {16}},\ \bibinfo {pages}
				{4776} (\bibinfo {year} {2020})}\BibitemShut {NoStop}%
			\bibitem [{\citenamefont {Santra}\ \emph {et~al.}(2020)\citenamefont {Santra},
				\citenamefont {Basu},\ and\ \citenamefont {Sabhapandit}}]{ion2020}%
			\BibitemOpen
			\bibfield  {author} {\bibinfo {author} {\bibfnamefont {I.}~\bibnamefont
					{Santra}}, \bibinfo {author} {\bibfnamefont {U.}~\bibnamefont {Basu}},\ and\
				\bibinfo {author} {\bibfnamefont {S.}~\bibnamefont {Sabhapandit}},\
			}\bibfield  {title} {\bibinfo {title} {Run-and-tumble particles in two
					dimensions: Marginal position distributions},\ }\href
			{https://doi.org/10.1103/PhysRevE.101.062120} {\bibfield  {journal} {\bibinfo
					{journal} {Phys. Rev. E}\ }\textbf {\bibinfo {volume} {101}},\ \bibinfo
				{pages} {062120} (\bibinfo {year} {2020})}\BibitemShut {NoStop}%
			\bibitem [{\citenamefont {Majumdar}\ and\ \citenamefont
				{Meerson}(2020)}]{Majumdar2020}%
			\BibitemOpen
			\bibfield  {author} {\bibinfo {author} {\bibfnamefont {S.~N.}\ \bibnamefont
					{Majumdar}}\ and\ \bibinfo {author} {\bibfnamefont {B.}~\bibnamefont
					{Meerson}},\ }\bibfield  {title} {\bibinfo {title} {{Toward the full
						short-time statistics of an active Brownian particle on the plane}},\ }\href
			{https://doi.org/10.1103/PhysRevE.102.022113} {\bibfield  {journal} {\bibinfo
					{journal} {Phys. Rev. E}\ }\textbf {\bibinfo {volume} {102}},\ \bibinfo
				{pages} {022113} (\bibinfo {year} {2020})}\BibitemShut {NoStop}%
			\bibitem [{\citenamefont {Malakar}\ \emph {et~al.}(2020)\citenamefont
				{Malakar}, \citenamefont {Das}, \citenamefont {Kundu}, \citenamefont
				{Kumar},\ and\ \citenamefont {Dhar}}]{Malakar2020}%
			\BibitemOpen
			\bibfield  {author} {\bibinfo {author} {\bibfnamefont {K.}~\bibnamefont
					{Malakar}}, \bibinfo {author} {\bibfnamefont {A.}~\bibnamefont {Das}},
				\bibinfo {author} {\bibfnamefont {A.}~\bibnamefont {Kundu}}, \bibinfo
				{author} {\bibfnamefont {K.~V.}\ \bibnamefont {Kumar}},\ and\ \bibinfo
				{author} {\bibfnamefont {A.}~\bibnamefont {Dhar}},\ }\bibfield  {title}
			{\bibinfo {title} {{Steady state of an active Brownian particle in a
						two-dimensional harmonic trap}},\ }\href
			{https://doi.org/10.1103/PhysRevE.101.022610} {\bibfield  {journal} {\bibinfo
					{journal} {Phys. Rev. E}\ }\textbf {\bibinfo {volume} {101}},\ \bibinfo
				{pages} {022610} (\bibinfo {year} {2020})}\BibitemShut {NoStop}%
			\bibitem [{\citenamefont {Basu}\ \emph {et~al.}(2020)\citenamefont {Basu},
				\citenamefont {Majumdar}, \citenamefont {Rosso}, \citenamefont
				{Sabhapandit},\ and\ \citenamefont {Schehr}}]{Basu2020}%
			\BibitemOpen
			\bibfield  {author} {\bibinfo {author} {\bibfnamefont {U.}~\bibnamefont
					{Basu}}, \bibinfo {author} {\bibfnamefont {S.~N.}\ \bibnamefont {Majumdar}},
				\bibinfo {author} {\bibfnamefont {A.}~\bibnamefont {Rosso}}, \bibinfo
				{author} {\bibfnamefont {S.}~\bibnamefont {Sabhapandit}},\ and\ \bibinfo
				{author} {\bibfnamefont {G.}~\bibnamefont {Schehr}},\ }\bibfield  {title}
			{\bibinfo {title} {{Exact stationary state of a run-and-tumble particle with
						three internal states in a harmonic trap}},\ }\bibfield  {journal} {\bibinfo
				{journal} {J. Phys. A: Math. Theor.}\ }\textbf {\bibinfo {volume} {53}},\
			\href {https://doi.org/10.1088/1751-8121/ab6af0} {10.1088/1751-8121/ab6af0}
			(\bibinfo {year} {2020})\BibitemShut {NoStop}%
			\bibitem [{\citenamefont {Santra}\ and\ \citenamefont {Basu}(2022)}]{ion2022}%
			\BibitemOpen
			\bibfield  {author} {\bibinfo {author} {\bibfnamefont {I.}~\bibnamefont
					{Santra}}\ and\ \bibinfo {author} {\bibfnamefont {U.}~\bibnamefont {Basu}},\
			}\bibfield  {title} {\bibinfo {title} {{Activity driven transport in harmonic
						chains}},\ }\href {https://doi.org/10.21468/SciPostPhys.13.2.041} {\bibfield
				{journal} {\bibinfo  {journal} {SciPost Phys.}\ }\textbf {\bibinfo {volume}
					{13}},\ \bibinfo {pages} {041} (\bibinfo {year} {2022})}\BibitemShut
			{NoStop}%
			\bibitem [{\citenamefont {Santra}\ \emph {et~al.}(2021)\citenamefont {Santra},
				\citenamefont {Basu},\ and\ \citenamefont {Sabhapandit}}]{Santra2021}%
			\BibitemOpen
			\bibfield  {author} {\bibinfo {author} {\bibfnamefont {I.}~\bibnamefont
					{Santra}}, \bibinfo {author} {\bibfnamefont {U.}~\bibnamefont {Basu}},\ and\
				\bibinfo {author} {\bibfnamefont {S.}~\bibnamefont {Sabhapandit}},\
			}\bibfield  {title} {\bibinfo {title} {{Active Brownian motion with
						directional reversals}},\ }\href
			{https://doi.org/10.1103/PhysRevE.104.L012601} {\bibfield  {journal}
				{\bibinfo  {journal} {Phys. Rev. E}\ }\textbf {\bibinfo {volume} {104}},\
				\bibinfo {pages} {L012601} (\bibinfo {year} {2021})}\BibitemShut {NoStop}%
			\bibitem [{\citenamefont {Chaudhuri}\ and\ \citenamefont
				{Dhar}(2021)}]{chaudhuri2021}%
			\BibitemOpen
			\bibfield  {author} {\bibinfo {author} {\bibfnamefont {D.}~\bibnamefont
					{Chaudhuri}}\ and\ \bibinfo {author} {\bibfnamefont {A.}~\bibnamefont
					{Dhar}},\ }\bibfield  {title} {\bibinfo {title} {Active {B}rownian particle
					in harmonic trap: {E}xact computation of moments, and re-entrant
					transition},\ }\href@noop {} {\bibfield  {journal} {\bibinfo  {journal} {J.
						Stat. Mech.: Theor. and Expt.}\ }\textbf {\bibinfo {volume} {2021}},\
				\bibinfo {pages} {013207} (\bibinfo {year} {2021})}\BibitemShut {NoStop}%
			\bibitem [{\citenamefont {Shee}\ and\ \citenamefont
				{Chaudhuri}(2022{\natexlab{a}})}]{Shee2022}%
			\BibitemOpen
			\bibfield  {author} {\bibinfo {author} {\bibfnamefont {A.}~\bibnamefont
					{Shee}}\ and\ \bibinfo {author} {\bibfnamefont {D.}~\bibnamefont
					{Chaudhuri}},\ }\bibfield  {title} {\bibinfo {title} {{Active Brownian motion
						with speed fluctuations in arbitrary dimensions: exact calculation of moments
						and dynamical crossovers}},\ }\href
			{https://doi.org/10.1088/1742-5468/ac403f} {\bibfield  {journal} {\bibinfo
					{journal} {J. Stat. Mech.: Theor. Expt.}\ }\textbf {\bibinfo {volume}
					{2022}},\ \bibinfo {pages} {013201} (\bibinfo {year}
				{2022}{\natexlab{a}})}\BibitemShut {NoStop}%
			\bibitem [{\citenamefont {Shee}\ and\ \citenamefont
				{Chaudhuri}(2022{\natexlab{b}})}]{Shee2022a}%
			\BibitemOpen
			\bibfield  {author} {\bibinfo {author} {\bibfnamefont {A.}~\bibnamefont
					{Shee}}\ and\ \bibinfo {author} {\bibfnamefont {D.}~\bibnamefont
					{Chaudhuri}},\ }\bibfield  {title} {\bibinfo {title} {{Self-propulsion with
						speed and orientation fluctuation: {E}xact computation of moments and
						dynamical bistabilities in displacement}},\ }\href
			{https://doi.org/10.1103/PhysRevE.105.054148} {\bibfield  {journal} {\bibinfo
					{journal} {Phys. Rev. E}\ }\textbf {\bibinfo {volume} {105}},\ \bibinfo
				{pages} {054148} (\bibinfo {year} {2022}{\natexlab{b}})}\BibitemShut
			{NoStop}%
			\bibitem [{\citenamefont {Dean}\ \emph {et~al.}(2021)\citenamefont {Dean},
				\citenamefont {Majumdar},\ and\ \citenamefont {Schawe}}]{Dean2021}%
			\BibitemOpen
			\bibfield  {author} {\bibinfo {author} {\bibfnamefont {D.~S.}\ \bibnamefont
					{Dean}}, \bibinfo {author} {\bibfnamefont {S.~N.}\ \bibnamefont {Majumdar}},\
				and\ \bibinfo {author} {\bibfnamefont {H.}~\bibnamefont {Schawe}},\
			}\bibfield  {title} {\bibinfo {title} {Position distribution in a generalized
					run-and-tumble process},\ }\href
			{https://doi.org/10.1103/PhysRevE.103.012130} {\bibfield  {journal} {\bibinfo
					{journal} {Phys. Rev. E}\ }\textbf {\bibinfo {volume} {103}},\ \bibinfo
				{pages} {012130} (\bibinfo {year} {2021})}\BibitemShut {NoStop}%
			\bibitem [{\citenamefont {Slowman}\ \emph {et~al.}(2017)\citenamefont
				{Slowman}, \citenamefont {Evans},\ and\ \citenamefont
				{Blythe}}]{slowman_2017}%
			\BibitemOpen
			\bibfield  {author} {\bibinfo {author} {\bibfnamefont {A.~B.}\ \bibnamefont
					{Slowman}}, \bibinfo {author} {\bibfnamefont {M.~R.}\ \bibnamefont {Evans}},\
				and\ \bibinfo {author} {\bibfnamefont {R.~A.}\ \bibnamefont {Blythe}},\
			}\bibfield  {title} {\bibinfo {title} {Exact solution of two interacting
					run-and-tumble random walkers with finite tumble duration},\ }\href
			{https://doi.org/10.1088/1751-8121/aa80af} {\bibfield  {journal} {\bibinfo
					{journal} {J. Phys. A: Math. and Theor.}\ }\textbf {\bibinfo {volume} {50}},\
				\bibinfo {pages} {375601} (\bibinfo {year} {2017})}\BibitemShut {NoStop}%
			\bibitem [{\citenamefont {Das}\ \emph {et~al.}(2020)\citenamefont {Das},
				\citenamefont {Dhar},\ and\ \citenamefont {Kundu}}]{das2020gap}%
			\BibitemOpen
			\bibfield  {author} {\bibinfo {author} {\bibfnamefont {A.}~\bibnamefont
					{Das}}, \bibinfo {author} {\bibfnamefont {A.}~\bibnamefont {Dhar}},\ and\
				\bibinfo {author} {\bibfnamefont {A.}~\bibnamefont {Kundu}},\ }\bibfield
			{title} {\bibinfo {title} {Gap statistics of two interacting run and tumble
					particles in one dimension},\ }\href@noop {} {\bibfield  {journal} {\bibinfo
					{journal} {J. Phys. A: Math. and Theor.}\ }\textbf {\bibinfo {volume} {53}},\
				\bibinfo {pages} {345003} (\bibinfo {year} {2020})}\BibitemShut {NoStop}%
			\bibitem [{\citenamefont {Le~Doussal}\ \emph {et~al.}(2019)\citenamefont
				{Le~Doussal}, \citenamefont {Majumdar},\ and\ \citenamefont
				{Schehr}}]{satya2019}%
			\BibitemOpen
			\bibfield  {author} {\bibinfo {author} {\bibfnamefont {P.}~\bibnamefont
					{Le~Doussal}}, \bibinfo {author} {\bibfnamefont {S.~N.}\ \bibnamefont
					{Majumdar}},\ and\ \bibinfo {author} {\bibfnamefont {G.}~\bibnamefont
					{Schehr}},\ }\bibfield  {title} {\bibinfo {title} {Noncrossing run-and-tumble
					particles on a line},\ }\href {https://doi.org/10.1103/PhysRevE.100.012113}
			{\bibfield  {journal} {\bibinfo  {journal} {Phys. Rev. E}\ }\textbf {\bibinfo
					{volume} {100}},\ \bibinfo {pages} {012113} (\bibinfo {year}
				{2019})}\BibitemShut {NoStop}%
			\bibitem [{\citenamefont {Slowman}\ \emph {et~al.}(2016)\citenamefont
				{Slowman}, \citenamefont {Evans},\ and\ \citenamefont
				{Blythe}}]{slowman2016}%
			\BibitemOpen
			\bibfield  {author} {\bibinfo {author} {\bibfnamefont {A.~B.}\ \bibnamefont
					{Slowman}}, \bibinfo {author} {\bibfnamefont {M.~R.}\ \bibnamefont {Evans}},\
				and\ \bibinfo {author} {\bibfnamefont {R.~A.}\ \bibnamefont {Blythe}},\
			}\bibfield  {title} {\bibinfo {title} {Jamming and attraction of interacting
					run-and-tumble random walkers},\ }\href
			{https://doi.org/10.1103/PhysRevLett.116.218101} {\bibfield  {journal}
				{\bibinfo  {journal} {Phys. Rev. Lett.}\ }\textbf {\bibinfo {volume} {116}},\
				\bibinfo {pages} {218101} (\bibinfo {year} {2016})}\BibitemShut {NoStop}%
			\bibitem [{\citenamefont {Kourbane-Houssene}\ \emph {et~al.}(2018)\citenamefont
				{Kourbane-Houssene}, \citenamefont {Erignoux}, \citenamefont {Bodineau},\
				and\ \citenamefont {Tailleur}}]{houssene2018}%
			\BibitemOpen
			\bibfield  {author} {\bibinfo {author} {\bibfnamefont {M.}~\bibnamefont
					{Kourbane-Houssene}}, \bibinfo {author} {\bibfnamefont {C.}~\bibnamefont
					{Erignoux}}, \bibinfo {author} {\bibfnamefont {T.}~\bibnamefont {Bodineau}},\
				and\ \bibinfo {author} {\bibfnamefont {J.}~\bibnamefont {Tailleur}},\
			}\bibfield  {title} {\bibinfo {title} {Exact hydrodynamic description of
					active lattice gases},\ }\href
			{https://doi.org/10.1103/PhysRevLett.120.268003} {\bibfield  {journal}
				{\bibinfo  {journal} {Phys. Rev. Lett.}\ }\textbf {\bibinfo {volume} {120}},\
				\bibinfo {pages} {268003} (\bibinfo {year} {2018})}\BibitemShut {NoStop}%
			\bibitem [{\citenamefont {Put}\ \emph {et~al.}(2019)\citenamefont {Put},
				\citenamefont {Berx},\ and\ \citenamefont {Vanderzande}}]{put2019}%
			\BibitemOpen
			\bibfield  {author} {\bibinfo {author} {\bibfnamefont {S.}~\bibnamefont
					{Put}}, \bibinfo {author} {\bibfnamefont {J.}~\bibnamefont {Berx}},\ and\
				\bibinfo {author} {\bibfnamefont {C.}~\bibnamefont {Vanderzande}},\
			}\bibfield  {title} {\bibinfo {title} {Non-{G}aussian anomalous dynamics in
					systems of interacting run-and-tumble particles},\ }\href
			{https://doi.org/10.1088/1742-5468/ab4e90} {\bibfield  {journal} {\bibinfo
					{journal} {J. Stat. Mech.: Theor. and Expt.}\ }\textbf {\bibinfo {volume}
					{2019}},\ \bibinfo {pages} {123205} (\bibinfo {year} {2019})}\BibitemShut
			{NoStop}%
			\bibitem [{\citenamefont {Dolai}\ \emph {et~al.}(2020)\citenamefont {Dolai},
				\citenamefont {Das}, \citenamefont {Kundu}, \citenamefont {Dasgupta},
				\citenamefont {Dhar},\ and\ \citenamefont {Kumar}}]{dolai_20}%
			\BibitemOpen
			\bibfield  {author} {\bibinfo {author} {\bibfnamefont {P.}~\bibnamefont
					{Dolai}}, \bibinfo {author} {\bibfnamefont {A.}~\bibnamefont {Das}}, \bibinfo
				{author} {\bibfnamefont {A.}~\bibnamefont {Kundu}}, \bibinfo {author}
				{\bibfnamefont {C.}~\bibnamefont {Dasgupta}}, \bibinfo {author}
				{\bibfnamefont {A.}~\bibnamefont {Dhar}},\ and\ \bibinfo {author}
				{\bibfnamefont {K.~V.}\ \bibnamefont {Kumar}},\ }\bibfield  {title} {\bibinfo
				{title} {Universal scaling in active single-file dynamics},\ }\href
			{https://doi.org/10.1039/D0SM00687D} {\bibfield  {journal} {\bibinfo
					{journal} {Soft Matter}\ }\textbf {\bibinfo {volume} {16}},\ \bibinfo {pages}
				{7077} (\bibinfo {year} {2020})}\BibitemShut {NoStop}%
			\bibitem [{\citenamefont {Singh}\ and\ \citenamefont
				{Kundu}(2021)}]{singh_2021}%
			\BibitemOpen
			\bibfield  {author} {\bibinfo {author} {\bibfnamefont {P.}~\bibnamefont
					{Singh}}\ and\ \bibinfo {author} {\bibfnamefont {A.}~\bibnamefont {Kundu}},\
			}\bibfield  {title} {\bibinfo {title} {Crossover behaviours exhibited by
					fluctuations and correlations in a chain of active particles},\ }\href
			{https://doi.org/10.1088/1751-8121/ac0a9f} {\bibfield  {journal} {\bibinfo
					{journal} {J. Phys. A: Math. and Theor.}\ }\textbf {\bibinfo {volume} {54}},\
				\bibinfo {pages} {305001} (\bibinfo {year} {2021})}\BibitemShut {NoStop}%
			\bibitem [{\citenamefont {Banerjee}\ \emph {et~al.}(2022)\citenamefont
				{Banerjee}, \citenamefont {Jack},\ and\ \citenamefont {Cates}}]{banerjee_22}%
			\BibitemOpen
			\bibfield  {author} {\bibinfo {author} {\bibfnamefont {T.}~\bibnamefont
					{Banerjee}}, \bibinfo {author} {\bibfnamefont {R.~L.}\ \bibnamefont {Jack}},\
				and\ \bibinfo {author} {\bibfnamefont {M.~E.}\ \bibnamefont {Cates}},\
			}\bibfield  {title} {\bibinfo {title} {Tracer dynamics in one dimensional
					gases of active or passive particles},\ }\href
			{https://doi.org/10.1088/1742-5468/ac4801} {\bibfield  {journal} {\bibinfo
					{journal} {J. Stat. Mech.: Theor. and Expt.}\ }\textbf {\bibinfo {volume}
					{2022}},\ \bibinfo {pages} {013209} (\bibinfo {year} {2022})}\BibitemShut
			{NoStop}%
			\bibitem [{\citenamefont {Touzo}\ \emph {et~al.}()\citenamefont {Touzo},
				\citenamefont {Doussal},\ and\ \citenamefont {Schehr}}]{touzo_2023}%
			\BibitemOpen
			\bibfield  {author} {\bibinfo {author} {\bibfnamefont {L.}~\bibnamefont
					{Touzo}}, \bibinfo {author} {\bibfnamefont {P.~L.}\ \bibnamefont {Doussal}},\
				and\ \bibinfo {author} {\bibfnamefont {G.}~\bibnamefont {Schehr}},\
			}\bibfield  {title} {\bibinfo {title} {Interacting, running and tumbling: The
					active {D}yson {B}rownian motion},\ }\href
			{https://doi.org/10.1209/0295-5075/acdabb} {\bibfield  {journal} {\bibinfo
					{journal} {Europhys. Lett.}\ }\textbf {\bibinfo {volume} {142}},\ \bibinfo
				{pages} {61004}}\BibitemShut {NoStop}%
			\bibitem [{\citenamefont {Agranov}\ \emph {et~al.}(2023)\citenamefont
				{Agranov}, \citenamefont {Ro}, \citenamefont {Kafri},\ and\ \citenamefont
				{Lecomte}}]{agranov2023}%
			\BibitemOpen
			\bibfield  {author} {\bibinfo {author} {\bibfnamefont {T.}~\bibnamefont
					{Agranov}}, \bibinfo {author} {\bibfnamefont {S.}~\bibnamefont {Ro}},
				\bibinfo {author} {\bibfnamefont {Y.}~\bibnamefont {Kafri}},\ and\ \bibinfo
				{author} {\bibfnamefont {V.}~\bibnamefont {Lecomte}},\ }\bibfield  {title}
			{\bibinfo {title} {{Macroscopic fluctuation theory and current fluctuations
						in active lattice gases}},\ }\href
			{https://doi.org/10.21468/SciPostPhys.14.3.045} {\bibfield  {journal}
				{\bibinfo  {journal} {SciPost Phys.}\ }\textbf {\bibinfo {volume} {14}},\
				\bibinfo {pages} {045} (\bibinfo {year} {2023})}\BibitemShut {NoStop}%
			\bibitem [{\citenamefont {Harris}(1965)}]{harris1965}%
			\BibitemOpen
			\bibfield  {author} {\bibinfo {author} {\bibfnamefont {T.~E.}\ \bibnamefont
					{Harris}},\ }\bibfield  {title} {\bibinfo {title} {Diffusion with
					“collisions” between particles},\ }\href@noop {} {\bibfield  {journal}
				{\bibinfo  {journal} {Journal of Applied Probability}\ }\textbf {\bibinfo
					{volume} {2}},\ \bibinfo {pages} {323} (\bibinfo {year} {1965})}\BibitemShut
			{NoStop}%
			\bibitem [{\citenamefont {Kollmann}(2003)}]{kollman2003}%
			\BibitemOpen
			\bibfield  {author} {\bibinfo {author} {\bibfnamefont {M.}~\bibnamefont
					{Kollmann}},\ }\bibfield  {title} {\bibinfo {title} {Single-file diffusion of
					atomic and colloidal systems: Asymptotic laws},\ }\href
			{https://doi.org/10.1103/PhysRevLett.90.180602} {\bibfield  {journal}
				{\bibinfo  {journal} {Phys. Rev. Lett.}\ }\textbf {\bibinfo {volume} {90}},\
				\bibinfo {pages} {180602} (\bibinfo {year} {2003})}\BibitemShut {NoStop}%
			\bibitem [{\citenamefont {Lizana}\ and\ \citenamefont
				{Ambj\"ornsson}(2008)}]{lizana2008}%
			\BibitemOpen
			\bibfield  {author} {\bibinfo {author} {\bibfnamefont {L.}~\bibnamefont
					{Lizana}}\ and\ \bibinfo {author} {\bibfnamefont {T.}~\bibnamefont
					{Ambj\"ornsson}},\ }\bibfield  {title} {\bibinfo {title} {Single-file
					diffusion in a box},\ }\href {https://doi.org/10.1103/PhysRevLett.100.200601}
			{\bibfield  {journal} {\bibinfo  {journal} {Phys. Rev. Lett.}\ }\textbf
				{\bibinfo {volume} {100}},\ \bibinfo {pages} {200601} (\bibinfo {year}
				{2008})}\BibitemShut {NoStop}%
			\bibitem [{\citenamefont {Hegde}\ \emph {et~al.}(2014)\citenamefont {Hegde},
				\citenamefont {Sabhapandit},\ and\ \citenamefont {Dhar}}]{hegde2014}%
			\BibitemOpen
			\bibfield  {author} {\bibinfo {author} {\bibfnamefont {C.}~\bibnamefont
					{Hegde}}, \bibinfo {author} {\bibfnamefont {S.}~\bibnamefont {Sabhapandit}},\
				and\ \bibinfo {author} {\bibfnamefont {A.}~\bibnamefont {Dhar}},\ }\bibfield
			{title} {\bibinfo {title} {Universal large deviations for the tagged particle
					in single-file motion},\ }\href
			{https://doi.org/10.1103/PhysRevLett.113.120601} {\bibfield  {journal}
				{\bibinfo  {journal} {Phys. Rev. Lett.}\ }\textbf {\bibinfo {volume} {113}},\
				\bibinfo {pages} {120601} (\bibinfo {year} {2014})}\BibitemShut {NoStop}%
			\bibitem [{\citenamefont {Krapivsky}\ \emph {et~al.}(2015)\citenamefont
				{Krapivsky}, \citenamefont {Mallick},\ and\ \citenamefont
				{Sadhu}}]{krapivsky2015}%
			\BibitemOpen
			\bibfield  {author} {\bibinfo {author} {\bibfnamefont {P.~L.}\ \bibnamefont
					{Krapivsky}}, \bibinfo {author} {\bibfnamefont {K.}~\bibnamefont {Mallick}},\
				and\ \bibinfo {author} {\bibfnamefont {T.}~\bibnamefont {Sadhu}},\ }\bibfield
			{title} {\bibinfo {title} {Tagged particle in single-file diffusion},\
			}\href@noop {} {\bibfield  {journal} {\bibinfo  {journal} {J. Stat. Phys.}\
				}\textbf {\bibinfo {volume} {160}},\ \bibinfo {pages} {885} (\bibinfo {year}
				{2015})}\BibitemShut {NoStop}%
			\bibitem [{\citenamefont {Wei}\ \emph {et~al.}(2000)\citenamefont {Wei},
				\citenamefont {Bechinger},\ and\ \citenamefont {Leiderer}}]{wei2000}%
			\BibitemOpen
			\bibfield  {author} {\bibinfo {author} {\bibfnamefont {Q.-H.}\ \bibnamefont
					{Wei}}, \bibinfo {author} {\bibfnamefont {C.}~\bibnamefont {Bechinger}},\
				and\ \bibinfo {author} {\bibfnamefont {P.}~\bibnamefont {Leiderer}},\
			}\bibfield  {title} {\bibinfo {title} {Single-file diffusion of colloids in
					one-dimensional channels},\ }\href@noop {} {\bibfield  {journal} {\bibinfo
					{journal} {Science}\ }\textbf {\bibinfo {volume} {287}},\ \bibinfo {pages}
				{625} (\bibinfo {year} {2000})}\BibitemShut {NoStop}%
			\bibitem [{\citenamefont {Lutz}\ \emph {et~al.}(2004)\citenamefont {Lutz},
				\citenamefont {Kollmann},\ and\ \citenamefont {Bechinger}}]{lutz2004}%
			\BibitemOpen
			\bibfield  {author} {\bibinfo {author} {\bibfnamefont {C.}~\bibnamefont
					{Lutz}}, \bibinfo {author} {\bibfnamefont {M.}~\bibnamefont {Kollmann}},\
				and\ \bibinfo {author} {\bibfnamefont {C.}~\bibnamefont {Bechinger}},\
			}\bibfield  {title} {\bibinfo {title} {Single-file diffusion of colloids in
					one-dimensional channels},\ }\href
			{https://doi.org/10.1103/PhysRevLett.93.026001} {\bibfield  {journal}
				{\bibinfo  {journal} {Phys. Rev. Lett.}\ }\textbf {\bibinfo {volume} {93}},\
				\bibinfo {pages} {026001} (\bibinfo {year} {2004})}\BibitemShut {NoStop}%
			\bibitem [{\citenamefont {Dhar}(2008)}]{dhar2008}%
			\BibitemOpen
			\bibfield  {author} {\bibinfo {author} {\bibfnamefont {A.}~\bibnamefont
					{Dhar}},\ }\bibfield  {title} {\bibinfo {title} {Heat transport in
					low-dimensional systems},\ }\href@noop {} {\bibfield  {journal} {\bibinfo
					{journal} {Advances in Physics}\ }\textbf {\bibinfo {volume} {57}},\ \bibinfo
				{pages} {457} (\bibinfo {year} {2008})}\BibitemShut {NoStop}%
			\bibitem [{\citenamefont {Hu}\ and\ \citenamefont {O'Connell}(1996)}]{hu_1996}%
			\BibitemOpen
			\bibfield  {author} {\bibinfo {author} {\bibfnamefont {G.~Y.}\ \bibnamefont
					{Hu}}\ and\ \bibinfo {author} {\bibfnamefont {R.~F.}\ \bibnamefont
					{O'Connell}},\ }\bibfield  {title} {\bibinfo {title} {Analytical inversion of
					symmetric tridiagonal matrices},\ }\href
			{https://doi.org/10.1088/0305-4470/29/7/020} {\bibfield  {journal} {\bibinfo
					{journal} {J. Phys. A: Math. Gen.}\ }\textbf {\bibinfo {volume} {29}},\
				\bibinfo {pages} {1511} (\bibinfo {year} {1996})}\BibitemShut {NoStop}%
			\bibitem [{\citenamefont {Lizana}\ \emph {et~al.}(2010)\citenamefont {Lizana},
				\citenamefont {Ambj{\"{o}}rnsson}, \citenamefont {Taloni}, \citenamefont
				{Barkai},\ and\ \citenamefont {Lomholt}}]{Lizana2010}%
			\BibitemOpen
			\bibfield  {author} {\bibinfo {author} {\bibfnamefont {L.}~\bibnamefont
					{Lizana}}, \bibinfo {author} {\bibfnamefont {T.}~\bibnamefont
					{Ambj{\"{o}}rnsson}}, \bibinfo {author} {\bibfnamefont {A.}~\bibnamefont
					{Taloni}}, \bibinfo {author} {\bibfnamefont {E.}~\bibnamefont {Barkai}},\
				and\ \bibinfo {author} {\bibfnamefont {M.~A.}\ \bibnamefont {Lomholt}},\
			}\bibfield  {title} {\bibinfo {title} {{Foundation of fractional Langevin
						equation: Harmonization of a many-body problem}},\ }\href
			{https://doi.org/10.1103/PhysRevE.81.051118} {\bibfield  {journal} {\bibinfo
					{journal} {Phys. Rev. E}\ }\textbf {\bibinfo {volume} {81}},\ \bibinfo
				{pages} {051118} (\bibinfo {year} {2010})}\BibitemShut {NoStop}%
			\bibitem [{\citenamefont {Dhar}(2001)}]{dhar2001}%
			\BibitemOpen
			\bibfield  {author} {\bibinfo {author} {\bibfnamefont {A.}~\bibnamefont
					{Dhar}},\ }\bibfield  {title} {\bibinfo {title} {Heat conduction in the
					disordered harmonic chain revisited},\ }\href
			{https://doi.org/10.1103/PhysRevLett.86.5882} {\bibfield  {journal} {\bibinfo
					{journal} {Phys. Rev. Lett.}\ }\textbf {\bibinfo {volume} {86}},\ \bibinfo
				{pages} {5882} (\bibinfo {year} {2001})}\BibitemShut {NoStop}%
			\bibitem [{\citenamefont {Dhar}\ \emph {et~al.}(2011)\citenamefont {Dhar},
				\citenamefont {Narayan}, \citenamefont {Kundu},\ and\ \citenamefont
				{Saito}}]{dhar_2011}%
			\BibitemOpen
			\bibfield  {author} {\bibinfo {author} {\bibfnamefont {A.}~\bibnamefont
					{Dhar}}, \bibinfo {author} {\bibfnamefont {O.}~\bibnamefont {Narayan}},
				\bibinfo {author} {\bibfnamefont {A.}~\bibnamefont {Kundu}},\ and\ \bibinfo
				{author} {\bibfnamefont {K.}~\bibnamefont {Saito}},\ }\bibfield  {title}
			{\bibinfo {title} {Linear-response formula for finite-frequency thermal
					conductance of open systems},\ }\href
			{https://doi.org/10.1103/PhysRevE.83.011101} {\bibfield  {journal} {\bibinfo
					{journal} {Phys. Rev. E}\ }\textbf {\bibinfo {volume} {83}},\ \bibinfo
				{pages} {011101} (\bibinfo {year} {2011})}\BibitemShut {NoStop}%
		\end{thebibliography}
		
		%
		
	\end{document}